\documentclass[12pt]{article}
\pdfoutput=1
\usepackage{color}
\usepackage{epsfig, palatino}
\usepackage{epic}
\usepackage{dsfont}
\usepackage{colonequals}
\usepackage{mathrsfs}
\usepackage{mdframed}
\usepackage{ae} 
\usepackage[T1]{fontenc}
\usepackage[ansinew]{inputenc}
\usepackage{amsmath}
\usepackage{amssymb}
\usepackage{graphicx}
\usepackage[normalem]{ulem}
\usepackage{xcolor}
\definecolor{darkblue}{cmyk}{0.9,0.9,0,0}
\usepackage[colorlinks=true,linkcolor=darkblue,citecolor=darkblue,urlcolor=darkblue]{hyperref}
\usepackage{cite}
\usepackage{hyperref}
\usepackage{inputenc}
\usepackage{varioref}
\usepackage{makeidx}
\usepackage[english]{babel}
\usepackage{simplewick}
\usepackage{array}
\usepackage{subcaption}
\usepackage{multirow}

\usepackage[font={small}]{caption}

\makeatletter

\input pdf-trans
\newbox\qbox
\def\usecolor#1{\csname\string\color@#1\endcsname\space}
\newcommand\bordercolor[1]{\colsplit{1}{#1}}
\newcommand\fillcolor[1]{\colsplit{0}{#1}}
\newcommand\outline[1]{\leavevmode%
  \def\maltext{#1}%
  \setbox\qbox=\hbox{\maltext}%
  \boxgs{Q q 2 Tr \thickness\space w \fillcol\space \bordercol\space}{}%
  \copy\qbox%
}
\makeatother
\newcommand\colsplit[2]{\colorlet{tmpcolor}{#2}\edef\tmp{\usecolor{tmpcolor}}%
  \def\tmpB{}\expandafter\colsplithelp\tmp\relax%
  \ifnum0=#1\relax\edef\fillcol{\tmpB}\else\edef\bordercol{\tmpC}\fi}
\def\colsplithelp#1#2 #3\relax{%
  \edef\tmpB{\tmpB#1#2 }%
  \ifnum `#1>`9\relax\def\tmpC{#3}\else\colsplithelp#3\relax\fi
}
\bordercolor{black}
\fillcolor{white}
\def\thickness{.3}

\definecolor{cadmiumgreen}{rgb}{0.0, 0.42, 0.24}

\newcommand{\comment}[1]{}

\newcommand{\begBvR}[1]{\begin{#1}} 
\newcommand{\beq}{\begBvR{equation}}
\newcommand{\eeq}{\end{equation}}

\newcommand{\eeqq}{\end{equation*}}

\newcommand\eeqaa{\end{eqnarray*}}

\newcommand\eeqa{\end{array}}

\newcommand{\eea}{\end{eqnarray}}

\newcommand{\im}{{\rm Im}\;}

\newcommand\IM{{\rm Im}\,}
\newcommand\RE{{\rm Re}\,}

\newcommand{\neqa}{\nonumber\end{eqnarray}} 
\newcommand{\la}[1]{\label{#1}}

\newcommand{\p}{\partial}

\renewcommand{\d}{\partial}

\newcommand{\<}{{\langle}}
\renewcommand{\>}{{\rangle}}

\newcommand{\re}{\relax{\rm I\kern-.18em R}}

\renewcommand{\sp}{p\hspace{-.40em}/}

\definecolor{darkgreen}{rgb}{0.0, 0.45, 0.0}
\definecolor{mathematicablue}{RGB}{94,130,182}

\def\XXint#1#2#3{{\setbox0=\hbox{$#1{#2#3}{\int}$}
\vcenter{\hbox{$#2#3$}}\kern-.5\wd0}}

\def\tr{{\rm tr~}}

\def\su2{{SU(2)}}

\def\[{\left[}
\def\]{\right]}

\def\({\left(}
\def\){\right)}
\def\[{\left[}
\def\]{\right]}

\def\<{\langle}
\def\>{\rangle}

\def\i2{\frac{i}{2}}

\def\spi{\relax{\rm \pi\kern-0.5em /}}
\def\sA{\relax{\rm A\kern-0.5em /}}
\def\sp{\relax{\rm p\kern-0.5em /}}
\def\sd{\relax{\rm \d\kern-0.5em /}}
\def\sk{\relax{\rm k\kern-0.5em /}}
\def\sn{\relax{\rm n\kern-0.5em /}}
\def\sl{\relax{\rm l\kern-0.5em /}}
\def\sP{\relax{\rm P\kern-0.7em /}}
\def\sBethe{\relax{\rm \Bethe\kern-0.5em /}}

\def\bbL{\text{\outline{$\Lambda$}}}
\def\bbr{\text{\outline{$\rho$}}}
\def\bbw{\text{\outline{w}}}

\def\2F1{\,_2{\rm F}_1}

\makeindex

\newcommand\blfootnote[1]{%
  \begingroup
  \renewcommand\thefootnote{}\footnote{\hspace{-6mm}#1}%
  \addtocounter{footnote}{-1}%
  \endgroup
}

\begin{document}

\thispagestyle{empty}

\renewcommand{\thefootnote}{\fnsymbol{footnote}}
\setcounter{page}{1}
\setcounter{footnote}{0}
\setcounter{figure}{0}

%\begin{flushright}
%CERN-TH-2017-162
%\end{flushright}
\vspace{-0.4in}

\begin{center}
$$$$
{\Large\textbf{\mathversion{bold}
Dual S-matrix Bootstrap I: 2D Theory}\par}
\vspace{1.0cm}

\textrm{Andrea L Guerrieri$^\text{\tiny 1}$, Alexandre Homrich$^\text{\tiny 1,\tiny 2,\tiny 3}$, Pedro Vieira$^\text{\tiny 1,\tiny 2}$}
\blfootnote{\tt  \#@gmail.com\&/@\{andrea.leonardo.guerrieri,alexandre.homrich,pedrogvieira\}}
\\ \vspace{1.2cm}
\footnotesize{\textit{
$^\text{\tiny 1}$ICTP South American Institute for Fundamental Research, IFT-UNESP, S\~ao Paulo, SP Brazil 01440-070   \\
$^\text{\tiny 2}$Perimeter Institute for Theoretical Physics,
Waterloo, Ontario N2L 2Y5, Canada\\
$^\text{\tiny 3}$Walter Burke Institute for Theoretical Physics, California Institute of Technology,
Pasadena, California 91125, USA\\
%$^\text{\tiny 3}$Fields and Strings Laboratory, 
%Institute of Physics, \'Ecole Polytechnique F\'ed\'erale de Lausanne (EPFL),
% CH-1015 Lausanne,
%Switzerland
%$^\text{\tiny 4}$Centre for Particle Theory, Department of Mathematical Sciences, Durham University, Lower Mountjoy, Stockton Road, Durham, England, DH1 3LE\\
}  
\vspace{4mm}
}
\end{center}

\par\vspace{1.5cm}

% \textbf{Abstract}

\vspace{2mm}
\begin{abstract}

Using duality in optimization theory we formulate a dual approach to the S-matrix bootstrap that provides rigorous bounds to 2D QFT observables as a consequence of unitarity, crossing symmetry and analyticity of the scattering matrix. We then explain how to optimize such bounds numerically, and prove that they provide the same bounds obtained from the usual primal formulation of the S-matrix Bootstrap, at least once convergence is attained from both perspectives. These techniques are then applied to the study of a gapped system with two stable particles of different masses, which serves as a toy model for bootstrapping popular physical systems.

\end{abstract}

\noindent

\setcounter{page}{1}
\renewcommand{\thefootnote}{\arabic{footnote}}
\setcounter{footnote}{0}

\setcounter{tocdepth}{2}

 \def\nref#1{{(\ref{#1})}}

\newpage

\tableofcontents

\parskip 5pt plus 1pt   \jot = 1.5ex

\newpage
\section{Introduction} 
\begin{figure}[h]
	\centering 
	\includegraphics[width=\linewidth]{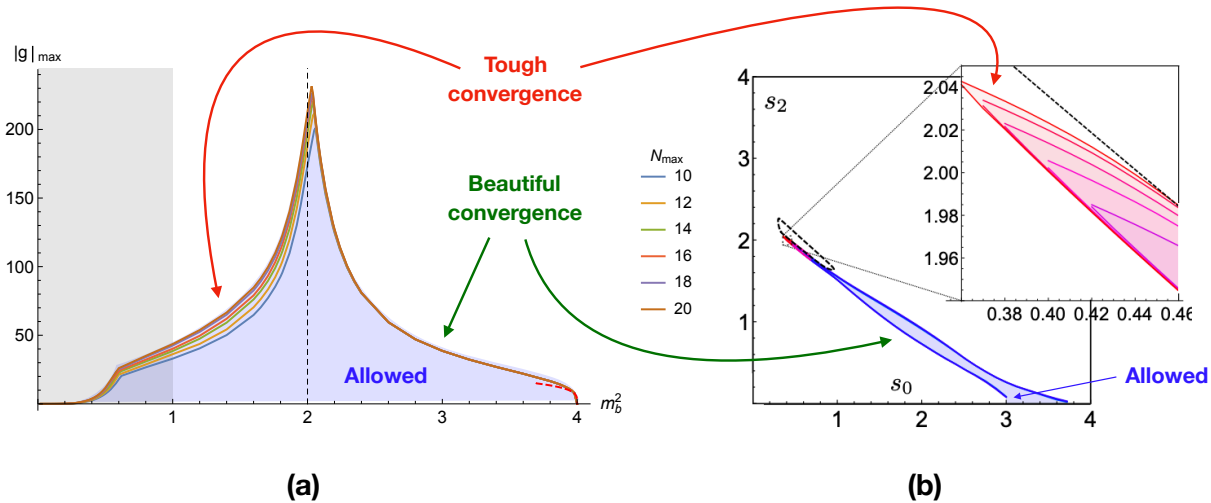}
	\vspace{-6.2cm}
	\caption{{\textbf a)} Maximal cubic coupling showing up in the scattering of the lightest particle in a gapped theory with a single bound-state (in this channel at least)~\cite{Paper3}. Convergence is perfect when the bound-state mass (measured in units of the lightest mass) is bigger than $\sqrt{2}$ and quite painful otherwise. {\textbf b)} The allowed chiral zeroes space of putative pion S-matrices associated to an $SU(2)$ chiral symmetry breaking patterns draws a beautiful peninsula like object with a sharp tip~\cite{Andrea}.\protect\footnotemark \,Convergence is great almost everywhere except close to the tip where numerics struggle. In those cases where the primal problem struggles, having a dual rigorous bound would be a blessing. This paper is about such dual bounds. }
	\label{figConvergenceintro}
\end{figure}

\la{intro}
Figure \ref{figConvergenceintro} is extracted from \cite{Paper3} and \cite{Andrea}. 
\footnotetext{There are, at least, other two structures would benefit a dual description. One is the ``pion lake''~\cite{Andrea}, found imposing the presence of the physical $\rho$ resonance only. Another interesting and recent structure is the ``pion river''~\cite{river}, found imposing additional constraints on the scattering lengths arising from $\chi$PT and monotonicity of the relative entropy. The dual formulation would allow to rigorously define these structures excluding theories not compatible with the assumed low energy QCD behavior.}

These works explore the allowed space of physical 4D S-matrices. One parametrizes a vast family of S-matrices compatible with given physical and mathematical assumptions and maximize or minimize quantities within this ansatz to find the boundaries of what is possible. The more parameters the ansatz has, the better is the exploration. As the number of parameters become very large, one hopes that these boundaries converge towards the true boundaries of the S-matrix space. 

Sometimes this works beautifully as illustrated in the figure; sometimes convergence is painful, to say the least, as also illustrated in the figure.
% \ref{figConvergence}. 
In those cases where convergence is a struggle, what can we do? Sometimes, it is a simple matter of improving the ansatz; sometimes it is not clear what exactly is missing. And in either case, how can we ever tell how close to converging are we  anyways?  

A solution would be to develop a dual numerical procedure -- called the \text{dual} problem -- where instead of constructing viable S-matrices we would instead rule out unphysical S-matrix space.\footnote{Such dual bounds were attempted more than 50 years ago already in \cite{Archeo1, Archeo2, Archeo3, Archeo4}. Would be very important to do some archeology work and revive/translate/re-discover/improve those old explorations in a modern computer friendly era. A beautiful first step is currently being pursued by Martin Kruczenski and Yifei He \cite{MartinTalkBootstrap}. The conformal bootstrap bounds are also exclusion analysis of this sort \cite{bootstrapReview}.} Then we would approach the boundaries of the S-matrix space from two sides, dual and primal, and in this way rigorously bracket the true boundaries of the sought after S-matrix space. 
This was recently achieved in two dimensions for simple models with a single type of particle transforming in some non-trivial global symmetry group \cite{Monolith}.\footnote{The primal version of these single particle studies with global symmetry was the subject of \cite{Martin,Miguel,Lucia}; the case without global symmetry was considered in \cite{Creutz,Paper2}.} 

This paper concerns two dimensional multi-particle systems with arbitrary mass spectra from this dual perspective, clearly one  step further in the complexity ladder, closer to the full higher dimensional problem.\footnote{Multi-particle primal problems of this kind were pioneered in \cite{Paper4,ToAppearSUSY}.} We will also consider a different technical approach, complementary to \cite{Monolith}, with some aspects which we hope can be more directly transposable to higher dimensions.

\section{Dual optimization and the S-matrix bootstrap}
\label{sec2}

\qquad To achieve the desired dual formulation, it is useful to revisit the S-matrix bootstrap with a slightly different perspective.

In the \textit{primal} S-matrix bootstrap formulation
% works so far (cite some people) were concerned in exploring what is possible in the space of physical S-matrices through the so called primal formulation. In it, 
 one constructs scattering amplitudes consistent with a set of axioms, or constraints. Such amplitudes are said to be \textit{feasible}, that is, they belong to the allowed space of theories. 
One then optimizes physical observables, such as the interaction strength between stable particles, in the space of feasible amplitudes. The prototypical example is \cite{Paper2,Creutz}: in a 2D theory with a single stable particle of mass $m$, what is the maximum cubic coupling $g$ consistent with a $2 \rightarrow 2$ scattering amplitude $M$ satisfying the constraints of unitarity, extended analyticity, and crossing? 

In other words, we would like to solve the optimization problem
         \begin{mdframed}[frametitle={Primal problem},frametitlealignment=\centering,backgroundcolor=blue!6, leftmargin=0cm, rightmargin=0cm, topline=false,
	bottomline=false, leftline=false, rightline=false] \label{primal}
	\vspace{-0.6cm}
  \begin{align} &\underset{\text{in } M(s)\text{, }g^2}{\text{maximize}} && g^2  \label{primal}\\
        & \text{constrained by} && \mathcal{A}(s) \equiv M(s) - \Big(M_\infty -\frac{g^2}{s-m^2} +\!\!\! \int\limits_{4m^2}^\infty \! \frac{dz}{\pi} \frac{\text{Im}M(z)}{s-z {+}i0} {+} \left(s \leftrightarrow 4m^2-s \right)\Big) =0 \nonumber\\
        & &&\text{for } s>4m^2, \label{analandcrossing}\\
        & \text{} && \mathcal{U}(s) \equiv 2\text{ Im}M(s) - \frac{\lvert M(s)\rvert^2}{2\sqrt{s-4m^2}\sqrt{s}} \geq  0 \qquad \text{for } s>4m^2. \label{unitarity}
       \end{align}
     \end{mdframed}
where we maximize over the space of analytic functions $M$, and emphasize that one parameter in this infinite dimensional space is the residue of such functions at $s=m^2$ which is equal to~$-g^2$. 
%Here, as usual, $s$ and $t$ correspond to the Mandelstam invariants\footnote{Recall that in 2D, $u=0$.}. 
The first constraint (\ref{analandcrossing}), an exact equality, imposes that feasible scattering amplitudes must respect crossing, real analyticity, and have singularities determined by physical processes: poles corresponding to one particle states, and cuts corresponding to multi-particle states.\footnote{It turns out that there is no loss of generality in omitting subtractions from (\ref{analandcrossing}), since a more careful analysis shows that the inclusion of those leads to the same result (\ref{dual bootstrap}). We opt for not including subtractions in the main text for the sake of clarity -- see appendix~\ref{analyticstuff} for a more detailed discussion.}
 We choose to impose this condition for $s > 4m^2$, but because we maximise over analytic functions, feasible amplitudes will have have this property for all $s$ in the physical sheet.\footnote{The physical sheet is defined as the first Riemann sheet encountered after analytically continuing from physical kinematics, $s>4m^2$, using the $+ i \epsilon$ prescription.
 } The convenience of imposing this condition for $s > 4m^2$ will become clear in time. The second constraint (\ref{unitarity}) is the physical unitarity condition, equivalent to~$\lvert S(s)\rvert \leq 1$. 

Since the quantity we are maximising, the objective, is a linear map in the space of analytic functions, the map that evaluates the residue at a point, and since the constraints~(\ref{analandcrossing}),~(\ref{unitarity}) are affine and convex respectively, the optimization problem we aim to solve is an infinite dimensional convex optimization problem. For such a simple problem, there are now two directions that can be taken. The first option is to solve the infinite dimensional problem analytically. As is well known by now, this follows from a simple application of the maximum modulus principle~\cite{Paper2, Creutz}. The second option, available in more complicated situations, is to bring the problem to the realm of computers by maximizing our objective in some finite dimensional subspace of analytic functions. For example, one can consider analytic functions that are, up to poles, polynomial of at most degree $N_\text{max}$ in some foliation variable $\rho$ that trivializes the constraint (\ref{analandcrossing}), as done in~\cite{Paper3}. This truncated problem can be efficiently solved by a convex optimization software, for example SDPB \cite{SDPB, scaling}. By choosing and increasing the finite dimensional subspace smartly, one obtains lower bounds to the solution of the primal problem that should converge to the correct bound with more expensive numerics.

The primal formulation suffers from two important shortcomings. First, for some problems it is hard to identify a simple ansatz, or truncation scheme, that allows for fast convergence. This is often the case in higher dimensional S-matrix bootstrap applications, or when scattering heavy particles in 2D. Second, and perhaps more importantly, one may want to add extra variables and constraints to the primal problem. In the previous example, those variables  and constraints could be, respectively, higher point amplitudes and higher point unitarity equations. It may be the case that a feasible $2 \rightarrow 2$ amplitude in the original primal problem may no longer be feasible in the enlarged space with extra constraints. In those cases, a point in theory space previously said to be allowed becomes forbidden. It would be more satisfying if bounds on the space of theories obtained by studying some scattering subsector remained true once the full set of QFT constraints were imposed.\footnote{Much in the same way that CFT data excluded by the numerical conformal bootstrap remains excluded once more crossing equations are included into the system.} To overcome both of this shortcomings, we introduce the dual formulation. We use the coupling maximization problem as a guiding example, before generalizing.

Consider the Lagrangian\footnote{Note $\mathcal{A}(s)$ is actually real. }
\beq
\mathcal{L}(M,w,\lambda) = g^2 + \int_{4 m^2}^\infty ds\text{ } w(s) \mathcal{A}(s) + \lambda(s) \mathcal{U}(s)  \label{lagrangian}
\eeq
with $\lambda(s) \geq 0$ and define the dual functional
\beq
d(w,\lambda) =  \underset{\{M, g\}}{\text{sup}} \mathcal{L}(M,w,\lambda)\label{dualfunctional}
\eeq
Notice that the supremum is taken over unconstrained analytic functions $M$.\footnote{It is useful to think of analytic functions as being defined through their independent real and imaginary parts along a line. Of course, if the dispersion (\ref{analandcrossing}) were to hold, then those would not be independent. However, since we maximise over generic analytic functions, we are free to treat $\text{Re }M$ and $\text{Im } M$  for $s>4m^2$ as independent.} The dual functional $d$ is the central object in the dual formulation due to the following property: 

  \begin{mdframed}[frametitle={Weak Duality},frametitlealignment=\centering,backgroundcolor=black!10, leftmargin=0cm, rightmargin=0cm, topline=false,
	bottomline=false, leftline=false, rightline=false] \label{duality}
	\beq
\text{Let the solution of the primal problem be $g_*^2$. Then	}
d(w,\lambda) \geq g_*^2. \label{weak}
\eeq
     \end{mdframed}
Weak duality holds due to two observations. First, note that since
\beq
\underset{\{\lambda \geq 0, w\}}{\text{inf  }} \mathcal{L}(M,w,\lambda) = 
\begin{cases}
    g^2& \text{if } M \text{ is feasible}\\
    -\infty              & \text{otherwise},
\end{cases}\label{since} 
\eeq
we have that 
\begin{equation*}
\normalfont g_*^2 = \underset{\{M, g\}}{\text{sup}} \left[ \underset{\{\lambda \geq 0, w\}}{\text{inf  }} \mathcal{L}(M,w,\lambda)\right].
\end{equation*}
Weak duality then follows from the max-min inequality
\beq
d(w,\lambda) \geq  \underset{\{\lambda \geq 0, w\}}{\text{inf  }} \left[ \underset{\{M, g\}}{\text{sup}} \mathcal{L}(M,w,\lambda) \right] \geq  \underset{\{M, g\}}{\text{sup}} \left[ \underset{\{\lambda \geq 0, w\}}{\text{inf  }} \mathcal{L}(M,w,\lambda)\right] = g_*^2. \label{maxmin}
\eeq

Exploring the $\{w,\lambda\}$ space, the space of dual variables, we therefore obtain upper bounds on the values of $g$ allowed by the axioms and exclude regions in theory space. This, in turn, partially solves the first shortcome of the primal formulation: by providing upper limits on the coupling, it bounds how far from converging an ineffective primal truncation scheme may be. To find the best possible upper bound, we solve the
                  \begin{mdframed}[frametitle={Dual problem (generic)},frametitlealignment=\centering,backgroundcolor=red!6, leftmargin=0cm, rightmargin=0cm, topline=false,
	bottomline=false, leftline=false, rightline=false] 
\vspace{-0.6cm}
  \begin{align} &\underset{\text{in } w(s)\text{, }\lambda(s)}{\text{minimize}} && d(w,\lambda) \label{dual generic}\\
        & \text{constrained by} && \lambda(s)\geq 0\nonumber
              \end{align}
     \end{mdframed}

The construction of dual functionals from a primal optimization problem is standard in optimization theory, but the particularities of the problems encountered in the S-matrix bootstrap lead to important simplifications. One of these is that the analyticity of the scattering amplitude is inherited by the dual variable $w(s)$, conjugate to the analyticity constraint. In fact, let's define a ``dual scattering function", $W(s)$\footnote{It is worth stressing that the introduction of an analytic function $W(s)$ is not mandatory. It is possible to work with real densities $w(s)$ and follow the argument presented in this section using the same logic. This possibility is particularly useful in higher dimensions if one wants to assume no more than the proven analyticity domains~\cite{Archeo4}.}, odd under crossing and whose absorptive part is $w(s)$: \beq
W(s) \equiv  \frac{1}{\pi}\int_{4m^2}^\infty dz \frac{w(z)}{s-z {+}i0} - \left(s \leftrightarrow 4m^2-s \right). \label{disp}
\eeq

Then, swapping a few integrals in (\ref{lagrangian}) and using $\frac{1}{\left(s-z{\pm}i0\right)} = \mp i \pi \delta(s-z) + \mathcal{P}\frac{1}{(s-z)}$ leads to a very simple representation for the lagrangian as
\beq
\mathcal{L}(M,W,\lambda) = g^2 \left(1 + \pi W(m^2)\right)  + \int_{4 m^2}^\infty ds\text{ } \text{Im}\left(W(s) M(s)\right) + \lambda(s)\, \mathcal{U}(s). 
\label{eq13}
\eeq
Note that the Lagrangian density is now manifestly local in $M$ as the Cauchy kernel from~(\ref{analandcrossing}) has been nicely absorbed into $W$. This locality, together with the quadratic nature of the constraint equations\footnote{Dispersions for higher point amplitudes are no longer expected to be quadratic in lower point functions due to the presence of Landau singularities.} leads to the next simplification over  generic dual optimization problems: we can perform both the maximization over $M$ in (\ref{dualfunctional}) and the minimization over~$\lambda$ in~\eqref{dual generic} exactly. We now analyze those in sequence. 

Before doing that, first notice, linearity of $\mathcal{L}$ under $g^2$ implies that 
\beq
d(W, \lambda) = + \infty \,\,\,\, \text{ unless } \,\,\,\,  \pi W(m^2)=-1. \label{normunless}
\eeq
This means that unless $W$ is properly normalized at $m^2$, the bounds obtained from the dual functional are vacuous. Hence, in solving the dual problem, there is no loss of generality in restricting ourselves to the space of $W$ satisfying the constrain in (\ref{normunless}).
 
 The linear Lagrange equations with respect to variations of $M(s)$ for $s>4m^2$ results in
 \beq
 M_\text{critical}(s) = \left[\text{Im}(W(s))/\lambda(s) + i \left(2\lambda(s)  + \text{Re}(W(s))/\lambda(s)\right)\right] /(2 \rho^2_{11}) .
\nonumber
 \eeq
where $ \rho^2_{11} = 1/(2 \sqrt{s-4m^2}\sqrt{s})$. Second order variations show that, indeed, this is a local maximum provided $\lambda(s)>0$. It follows from the definition (\ref{dualfunctional}) that, provided $\pi W(m^2)=-1$,

\beq
d(W, \lambda) = \int_{4m^2}^\infty ds    \left( \frac{\lvert W(s)\rvert^2}{4 \lambda(s)} + \lambda(s) + \text{Re}W(s))\right)/ \rho^2_{11} . \la{dWL}
\eeq

Next, we minimize over $\lambda$ leading to $\lambda=|W(s)|/2$. The result is $D(W) \equiv \underset{\lambda \geq 0}{\text{inf  }} d(W,\lambda))$ given by 
\beq
D(W) = \int_{4m^2}^\infty ds  \left(\text{Re}(W(s)) + \lvert W(s)\rvert \right)/  \rho^2_{11}. \label{bfunc} ,
\eeq
in which case\footnote{Note that unitarity is automatically saturated once we minimize in $\lambda$.} 
\beq
 M_{\text{critical}}(s) = \frac{i}{\rho^2_{11}}\left(1 + \frac{W^*}{\lvert W\rvert} \right).\nonumber
\eeq

In sum, the dual of (\ref{primal}) simplifies to

                  \begin{mdframed}[frametitle={Dual problem (S-matrix bootstrap)},frametitlealignment=\centering,backgroundcolor=red!6, leftmargin=0cm, rightmargin=0cm, topline=false,
	bottomline=false, leftline=false, rightline=false] 
	\vspace{-0.4cm}
  \begin{align} &\underset{\text{in } W(s)}{\text{minimize}} && D(W)=\int_{4m^2}^\infty ds  \left(\text{Re}(W(s)) + \lvert W(s)\rvert \right)/  \rho^2_{11} \label{dual bootstrap}\\
        & \text{constrained by} && \pi W(m^2)=-1.  \label{norm}
      \end{align}
     \end{mdframed}

The dual problem can be tackled numerically through the same strategy used for the primal problem, that is, restricting our search to a finite dimensional subspace of analytic $W$s. For example, one could use the $\rho$ foliation variables to write the ansatz\footnote{The Ansatz (\ref{wansatz}) is consistent with the dispersion (\ref{disp}). In particular, the poles in (\ref{wansatz}) correspond to a delta function contribution in $w(s)$.}
\beq
W_{\text{ansatz}}(s) = \frac{1}{s (4m^2-s)}\sum_{n=1}^{N_\text{max}}a_n (\rho(s)^n - \rho(t)^n), \label{wansatz}
\eeq
where 
\beq
\rho(s) = \frac{\sqrt{2m^2 } - \sqrt{4m^2 -s}}{\sqrt{2m^2} + \sqrt{4m^2 -s}},
\label{rhovariabledef}
\eeq
and minimize the functional (\ref{dual bootstrap}) in the finite dimensional space parametrized by the $a_n$'s. Note that the constraint (\ref{norm}) is a linear constraint in this space. The functional (\ref{bfunc}) is nonlinear, but it is convex in $W$. Performing such minimization, say, in \texttt{Mathematica} shows that, as one increases $N_\text{max}$, the result of the problem (\ref{dual bootstrap}) converges to the result of the primal problem (\ref{primal}). This is expected if our optimization problem satisfies

%           black!10       
\begin{mdframed}[frametitle={Strong Duality},frametitlealignment=\centering,backgroundcolor=black!10, leftmargin=0cm, rightmargin=0cm, topline=false,
	bottomline=false, leftline=false, rightline=false]
The solutions to the primal (\ref{primal}) and dual problem (\ref{dual bootstrap}) are identical, i.e. $g_*^2 = \underset{\text{in } W}{\text{min}} \text{   } D(W).$ In other words, the $\ge$ symbol in (\ref{weak}) is actually an $=$ sign.
     \end{mdframed}
This property is argued for in appendix \ref{strong}.
 
To explain how the dual formulation solves the second shortcoming of the primal optimization, and in view of the applications in section \ref{application}, let's consider a slightly different class of S-matrix Bootstrap problems. Consider a gapped theory with two real stable particles of masses $m_1$ and $m_2$ respectively, $m_1<m_2$, and suppose we were interested in maximizing the cubic coupling of particle $m_1$. Let $\mathbb{M}_{ab} = M_{a\to b}$. Assuming $P$ and $T$ symmetry, $\mathbb{M}$ is a symmetric matrix. We would like to solve the problem
                  \begin{mdframed}[frametitle={Primal problem (matrix)},frametitlealignment=\centering,backgroundcolor=blue!6, leftmargin=0cm, rightmargin=0cm, topline=false,
	bottomline=false, leftline=false, rightline=false]
	\vspace{-0.6cm}
  \begin{align} &\underset{\text{in } \mathbb{M}}{\text{maximize}} && g^2  \label{primal2}\\
        & \text{constrained by} && \mathbb{A}(s) =0 &&&\text{for } s>4m_1^2, \label{analandcrossing2}\\
        & \text{} && \mathbb{U}(s) \equiv 2\text{ Im}\,\mathbb{M}(s) - \mathbb{M}^\dagger \bbr\, \mathbb{M} \succeq 0 \qquad \qquad &&&\text{for } s>4m_1^2. \label{unitarity2}
       \end{align}
     \end{mdframed}
where $\mathbb{A}_{a b} \equiv \mathcal{A}_{a \to b}$ are analogous to (\ref{analandcrossing}) and impose the correct dispersion relations for the amplitudes $M_{a \to b}$ (see e.g. (\ref{AA}) in the next section). Here $\bbr$ are the phase space factors for the intermediate states (see e.g. (\ref{rhoMatrix}) in the next section). 
%In the case of two particle of mass $m_c$,$m_d$, $\rho_{c d} = \frac{\theta\left(s-(m_c + m_d)^2\right)}{2\sqrt{s-(m_c + m_d)^2}\sqrt{s-(m_c - m_d)^2}}$. 
To obtain the dual problem, we introduce the Lagrangian
\beq
\mathcal{L}(\mathbb{M}, \bbw ,\bbL) = g^2 + \int_{4 m_1^2}^\infty ds\text{ } \text{Tr}\left( \bbw \cdot  \mathbb{A} (s) + \bbL \cdot \mathbb{U} (s) \right), \label{lagrangian2}
\eeq
where  $\bbw$ and $\bbL$ are respectively symmetric and hermitian matrices of dual variables with~$\text{\outline{$\Lambda$}}$ positive semi-definite. The new dual functional
\beq
d(\bbw, \bbL) =  \underset{\mathbb{M}}{\text{sup  }} \mathcal{L}(\mathbb{M}, \bbw ,{\bbL}) \label{dualfunctional2}
\eeq
satisfies weak duality by similar arguments as those in equations (\ref{since}-\ref{maxmin}). The dual optimization problem is 
   \begin{mdframed}[frametitle={Dual problem (matrix)},frametitlealignment=\centering,backgroundcolor=red!6, leftmargin=0cm, rightmargin=0cm, topline=false,
	bottomline=false, leftline=false, rightline=false] 
	\vspace{-0.6cm}
  \begin{align} &\underset{\text{in } \bbw(s)\text{, }\bbL(s)}{\text{minimize}} && d(\bbw, \bbL) \label{dual 2}\\
        & \text{constrained by} && \bbL(s)\succeq 0. \nonumber
      \end{align}
     \end{mdframed}
 Note that an upper bound on the solution of the primal problem (\ref{primal}) is obtained by  choosing minimizing $d$ in the subspace $\bbw_{ab}(s) = \delta_a^{11} \delta_b^{11} w(s)$, $\bbL_{ab} =  \delta_a^{11} \delta_b^{11} \lambda(s)$, $\lambda\geq0$. This is equivalent to the dual problem obtained by including only the amplitude $M_{11\to11}$ in the bootstrap system, or primal problem.  Restricting to a scattering subsector in the dual formulation provides true bounds to the more complete optimization problem. Conversely, bounds obtained by studying some restricted space of amplitudes and constrains remain valid once extra axioms and degrees of freedom are considered. We hope it is clear that the argument provided by means of an example is generic. This solves the second shortcoming of the primal formulation.

\section{An application}
\label{application}
\subsection{The setup}
We now turn our attention to much richer S-matrix bootstrap. We consider a theory with two particles of mass $m_1$ and $m_2>m_1$. We will \textit{not} assume any global symmetry.  For concreteness, we will take\footnote{Setting $m_1=1$ simply sets our units. All $m_2> \sqrt{2}$ would then give very similar plots/conclusions. We could also consider $m_2<\sqrt{2}$; the plots are a little bit less eye pleasing in that case. The significance of the transition point $m_2^*=\sqrt{2}$ is that this is the crossing invariant point for the $11\to 11$ process; on either sign of this point residues have different signs leading to quite different optimization results.}
\beq
m_1=1\,, \qquad m_2=3/2 \,.\nonumber
\eeq
There are a priori four couplings involving these two particles: $g_{111},g_{112},g_{122},g_{222}$. They would show up as $s$-channel residues in the various scattering amplitudes:
\beq
\begin{array}{c|c|c}
\text{Amplitude} & \text{Exchange of particle } 1 & \text{Exchange of particle } 2  \\  \hline
11\to 11 &  {\color{red}g_{111}^2}  &  {\color{blue} g_{112}^2} \\  \hline
11\to 12 &{\color{red}g_{111}}{\color{blue} g_{112}}  &{\color{blue} g_{112}}{\color{cadmiumgreen}g_{122}} \\  \hline
12\to 12 & {\color{blue} g_{112}^2} & {\color{cadmiumgreen} g_{122}^2} \\ \hline
11\to 22 & {\color{red} g_{111}} {\color{cadmiumgreen} g_{122}}  & {\color{blue} g_{112}} {\color{magenta}g_{222}} \\ \hline
12\to 22 & {\color{blue} g_{112}}{\color{cadmiumgreen} g_{122}} & {\color{cadmiumgreen} g_{122}} {\color{magenta} g_{222}} \\ \hline
22\to 22 & {\color{cadmiumgreen} g_{122}^2}  & {\color{magenta} g_{222}^2 }
\end{array} \nonumber
\eeq
We will not consider the full coupled system of six amplitudes. Instead we will consider a nice closed subset involving the $11\to 11$, $11\to 12$ and (the forward) $12\to 12$ processes only (that is, the first three lines in the table). As such we will be insensitive to $g_{222}$. We will furthermore consider a section of the remaining three-dimensional space where $g_{122}=0$ so that the problem simplifies slightly to\footnote{The analysis for any other fixed value of $g_{122}$ follows identically, see more at the end of this section. }
\beq
\begin{array}{c|c|c}
\text{Amplitude} & \text{Exchange of particle } 1 & \text{Exchange of particle } 2  \\  \hline
11\to 11 & {\color{red}g_{111}^2} & {\color{blue} g_{112}^2} \\  \hline
11\to 12 &{\color{red}g_{111}}{\color{blue} g_{112}} & 0 \\  \hline
12\to 12 & {\color{blue} g_{112}^2} & 0 
\end{array} \nonumber
\eeq
and our main goal here is to explore the allowed two dimensional $(g_{112},g_{111})$ space. A convenient way to find the boundary of this space is by shooting radially. We fix an angle $\beta$ and define a radius $R$ as
\beq
(g_{112},g_{111}) = R(\cos\beta,\sin\beta) \,. \nonumber
\eeq
Then we find the maximum value of $R$ for each $\beta$ choice to plot the full two-dimensional space. 

In the primal language we will get larger and larger $R$'s as our ansatz is more and more complete. In the dual language we will rule out smaller and smaller $R$ as we improve our ansatz. Sandwiched between the two will be the true (two dimensional section of the) boundary of the S-matrix space. 

It is equally straightforward to fix $g_{122}$ to any other value and analyze another 2d section in this way or even collect various values of $g_{122}$ to construct the full $3D$ space. We leave such detailed scans for the future when we will have more realistic setups designed to bootstrap particular relevant physical theories such as the (regular and tricritical) Ising model (perturbed by thermal and magnetic deformations) as discussed in the conclusions. 

\subsection{Single Component Horn}
\label{Horn}
Let us start our search for the two dimensional section of the allowed S-matrix space by focusing on the constraints arising from the single $M=M_{11\to 11}$ component alone. 

\begin{figure}[t]
	\centering 
	\includegraphics[width=\linewidth]{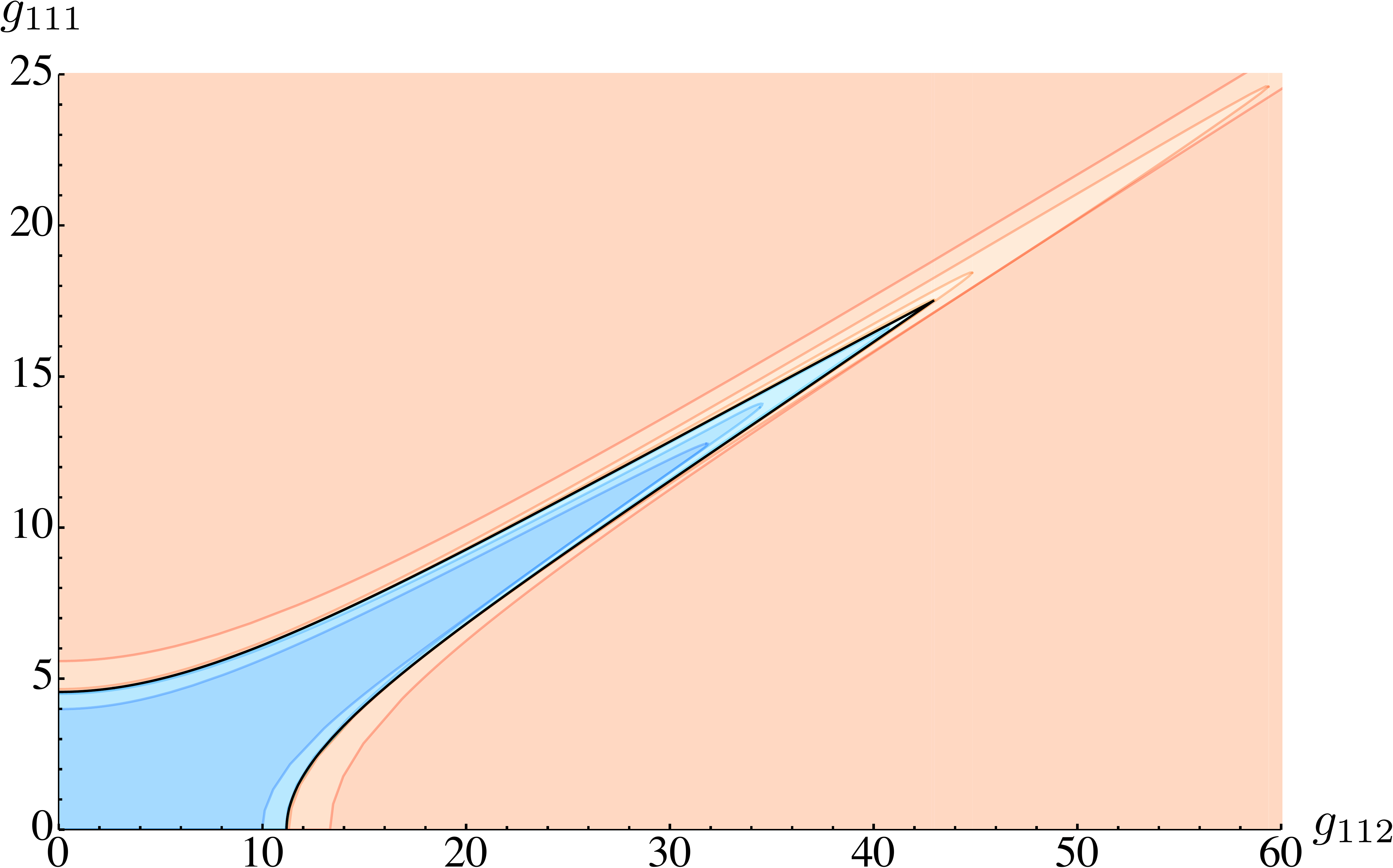}
	\caption{Numerical bounds on the coupling space $\{g_{111},g_{112}\}$. The blue shaded regions enclose the allowed points for different $N_{\text{max}}$ in our primal ansatz. The red shaded regions mark the points that are rigorously excluded. The thin black analytic curve is the boundary of the allowed region \cite{Paper4}. 
	 As we increase $N_{\max}$ from 1 to 5 in the primal problem, the blue regions enlarge, allowing for more and more points and eventually converging to touch the boundary of the permitted space (this is more evident in the ``horn'' region). In the dual strategy as we increase $N_{\max}$ from 1 to 5 we exclude more and more points. At convergence the excluded region touches the boundary of the allowed space. We restrict the plot to the first quadrant since it is symmetric under $g \leftrightarrow -g$.}
	\label{figHorn}
\end{figure}

This is a warm up section and many of the results here are not new: indeed, the primal formulation of single component scattering has been the subject of \cite{Paper2}; a minor new ingredient we will consider here is the radial search element. (The radial problem for the space of S-matrices with $O(N)$ symmetry and no bound states was introduced  in~\cite{Monolith}.) In appendix H of \cite{Paper4} an almost identical primal problem was solved analytically; the analytic curves in figure \ref{figHorn} are obtained by trivially adapting the arguments therein. The dual formulation for these single component cases with several exchanges masses, however, will be novel and provide very useful intuition for the most general case.  

%\red{way too fast. reader desn't even know we are talking about 11->11 (define $M = M1111$). need an introductory paragraph. Also, in general, for the whole of section 3 we should spell what is the outline for it, otherwise the reader is just thrown information without knowing where we are going with it. also maybe add titles to the boxes. e.g. the first "radial primal problem".  I already got confused with what "original  radial problem" refered to downstairs. title would help.}
The primal radial problem can be compactly formulated as 

\begin{mdframed}[frametitle={Primal Radial Problem for Single Component},frametitlealignment=\centering,backgroundcolor=blue!6, leftmargin=0cm, rightmargin=0cm, topline=false,bottomline=false, leftline=false, rightline=false] 
\vspace{-0.6cm}
\begin{align} &\underset{\text{in } {M, R^2}}{\text{maximize}} &&  R^2\nonumber\\
& \text{constr. by}  && \text{Res}_{m_1^2}(M)=R^2\sin^2\beta, \quad \text{Res}_{m_2^2}(M)=R^2\cos^2\beta \label{radialcondition}\\ 
& s\geq4m_1^2  && \mathcal{A}(s)=M(s){-}M_\infty+\left(\frac{g_{111}^2}{s-m_1^2}{+}\frac{g_{112}^2}{s-m_2^2}{-}\frac{1}{\pi}\int_{4m_1^2}^\infty dz\,\frac{\IM M(z)}{z-s} +(s\leftrightarrow t)\right){=}0\nonumber\\
& s\geq 4m_1^2&& \mathcal{U}(s)=2\IM M(s) -\rho_{11}^2 |M(s)|^2 \geq 0. 
 \label{primal bootstrap 11to11}
\end{align}
\end{mdframed}

We will now construct the dual problem. If it were not for the radial additional equality constraints~\eqref{radialcondition} the corresponding dual problem would be given already in eq.~\eqref{dual bootstrap}.
In this case we need to introduce additional Lagrange multipliers $\nu_1$ and $\nu_2$
to the lagrangian~\eqref{lagrangian}
\beq
\mathcal{L}=R^2+\nu_1 (\text{Res}_{m_1^2}(M)-R^2\sin^2\beta)+\nu_2 (\text{Res}_{m_2^2}(M)-R^2\cos^2\beta)+\int_{4m_1^2}^\infty ds\,\mathcal{A}(s)w(s)+\mathcal{U}(s)\lambda(s).
\label{lagrangianhorn}
\eeq
Now we follow the logic of section \ref{sec2} verbatin modulo a few small differences inherent to the radial nature of the primal problem which we will highlight. First of all note that the maximum of the Lagrangian with respect to $R^2$ yields a bounded result only when
\beq
1-\nu_1 \sin^2\beta-\nu_2 \cos^2\beta=0.\nonumber
\eeq
Next, identifying $w(s)=\IM W(s)$ with $W(s)$ given by (\ref{disp}) as before will lead to a beautiful dual problem formulation with a totally local optimization target. Importantly 
\beq
\int_{4m_1^2}^\infty ds\,\mathcal{A}(s)w(s)= \int_{4m_1^2}^\infty ds\,\text{Im}(M(s)W(s))+ \pi \text{Res}_{m_1^2}(M) W(m_1^2)+\pi  \text{Res}_{m_2^2}(M) W(m_2^2)\nonumber
\eeq
so we see that the optimization with respect to the parameters $ \text{Res}_{m_i^2}(M) $ identifies the lagrange multipliers $\nu_i$ with the normalization of the dual functional at the stable mass values $W(m_i^2)$. All in all we therefore obtain the simple dual problem radial generalization of (\ref{dual bootstrap}) as 
\begin{mdframed}[frametitle={Dual Radial Problem for Single Component},frametitlealignment=\centering,backgroundcolor=red!6, leftmargin=0cm, rightmargin=0cm, topline=false,bottomline=false, leftline=false, rightline=false] 
\vspace{-0.3cm}
\begin{align} &\underset{\text{in } {W}}{\text{minimize}} &&  D(W)=\int_{4m^2_1}^\infty ds  \left(\text{Re}(W(s)) + \lvert W(s)\rvert \right)/  \rho^2_{11}\nonumber\\
& \text{constrained by}  && 1+\pi\, W(m_1^2)\sin^2\beta+\pi\, W(m_2^2)\cos^2\beta=0.
\label{dual bootstrap 11to11}
\end{align}
\end{mdframed}

Notice again the nice complementarity between the pole singularities associated to bound states in the physical amplitude and the absence of poles in the ``dual scattering function" $W$ given by (\ref{disp}), replaced instead by the simple normalization conditions (\ref{dual bootstrap 11to11}). Conversely, when we maximize effective couplings in theories without bound-states the primal S-matrices have no bound-states and the dual functionals have poles \cite{Monolith}. 

In figure~\ref{figHorn} we show the numerical results for both the primal (inner blue shaded regions) and the dual problem (outer red shaded regions).

%\red{Again, reminding the reader what we are doing (here or in the intro paragraph mentioned above) is good I think. ``Next, we maximize with respect to primal variables R,M, to obtain the dual functional $d(W, \lambda,\nu)$". We were confused about this max vs min for a long time, the reader may already have forgotten by now.}  {\color{magenta}P: I stressed we are copying section 2 so the reader can go there to refresh memory. Enough?}

\subsection{Multiple Component Kinematics}

Next we consider the full system with $11\to 11$, $11\to 12$ and \textit{forward} $12\to 12$ amplitudes.\footnote{As reviewed in detail in \cite{Paper4} when a particle of type $1$ scatters with a particle of type $2$ it can either continue straight (\textit{forward amplitude}) or bounce back (\textit{backward amplitude}). Here we consider the forward process only. This process is nicely crossing symmetric. (The backward process is not; instead it is related by crossing to $11\to 22$ scattering so considering this backward process would require more scattering processes to close the system of {unitarity} equations.)} The two dimensional kinematics of the $11\to 11$ process and of the \textit{forward} $12\to 12$ process are reviewed in great detail in section 2 of \cite{Paper4} so here we will mostly focus on the new $11\to 12$ process.\footnote{This process was not considered in \cite{Paper4} because it violates $\mathbb{Z}_2$ symmetry. Here we don't have $\mathbb{Z}_2$ symmetry so it is the first most natural process to consider after the lightest $11\to 11$ scattering amplitude.} 
This scattering process is a nice fully symmetric process. No matter which channel we look at it, it always describes two particles of type $1$ (in the infinite future or past) scattering into a particle of type $1$ and another of type $2$. As such 
\beq
M_{11\to 12}(s,t,u)\nonumber
\eeq
is fully symmetric under any permutation of the three Mandelstam variables $s,t,u$. Of course, they are not independent. Besides 
\beq
s+t+u=3m_1^2+m_2^2 \la{Plus}
\eeq 
which holds in any dimension, we have the two dimensional constraint 
\beq
s t u=m_1^2\left(m_1^2- m_2^2\right){}^2 \la{Times}
\eeq

\begin{figure}[t]
	\centering 
	\includegraphics[width=\linewidth]{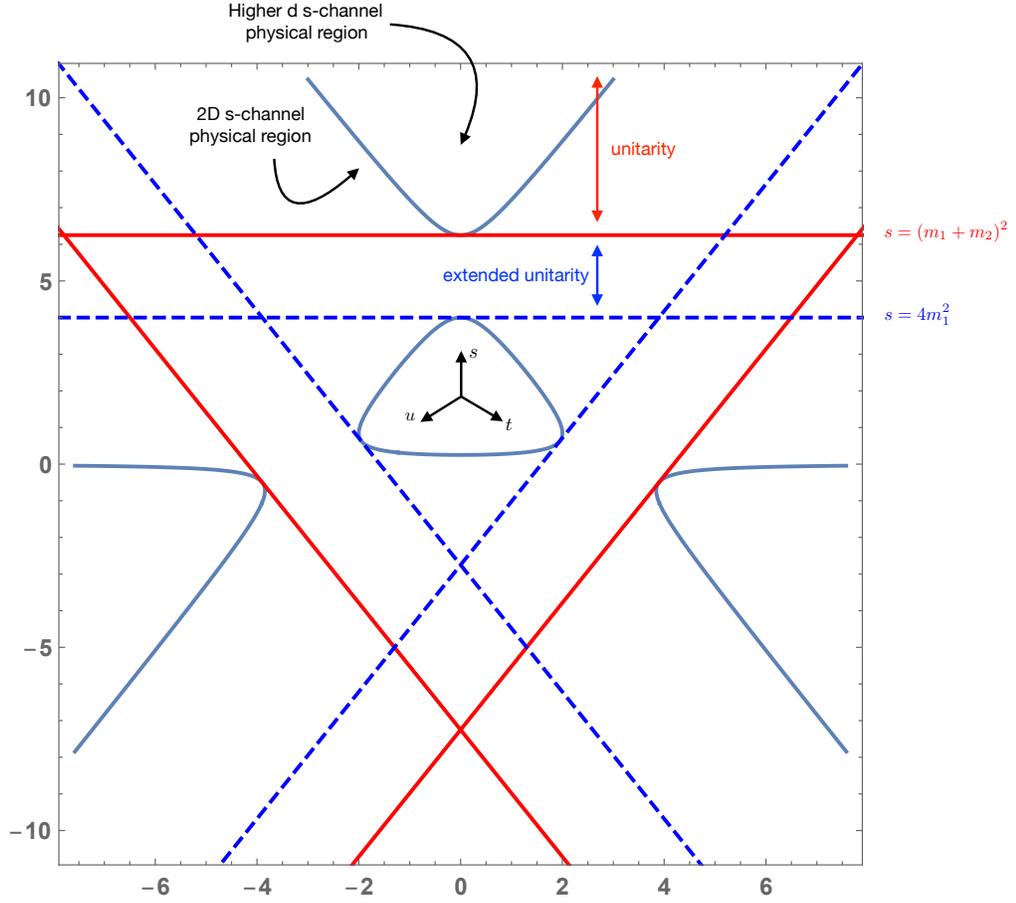}
%	\vspace{-6cm}
	\caption{Maldelstam Triangle for $11\to 12$ scattering. The x-axis is given by $x=(s+2 t-3 m_1^2-m_2^2)/\sqrt{3}$. The $11\to 12$ scattering if fully crossing invariant and indeed so is this picture. Physical processes in 2D lie on top of the blue solid lines and outside the red lines; in higher dimensions they fill in the interior of the regions delimited by the blue solid lines as one scans over physical scattering angles. Similar triangle for $12\to 12$ scattering can be found in \cite{Paper4}.}
	\label{triangle}
\end{figure}

Equations (\ref{Plus}) and (\ref{Times}) describe a curve. Its projection into real $s,t,u$ is given by the solid curved blue lines in figure \ref{triangle}. 
There, we see four disconnected regions: three non-compact parabola like curves related by a rotation symmetry and a round triangle in the middle. The three outer curves are the three physical regions associated to the three scattering channels. The one in the top, for instance, corresponds to the $s$-channel. (Each outer curve has a left and right components which are equivalent; they are related to a simple parity transformation.) 
The $s$-channel outer curve start at $s=(m_1+m_2)^2$ as indicated by the red solid line. That corresponds to the minimal energy necessary to produce a particle of type $1$ and a particle of mass $2$ at rest. (Recall that $2$ is heavier than $1$.) Another important energy marked by the blue dashed line in the figure occurs at $s=(2m_1)^2$ which would correspond to the minimal energy necessary to produce two particle of type $1$ at rest. This is however \textit{not} a physical energy for this process since physical energies are those for which we can produce \textit{both} initial \textit{and} final state. Nonetheless, the region between $s=4m_1^2$ and $s=(m_1+m_2)^2$ is very interesting because we know precisely what are the only possible {physical} states in that energy range: they can only be two particle states involving two particles of type $1$.~\cite{Landau} The equation which reflects this is the so called \textit{extended} unitarity relation which in this case reads
\beq
2\IM M_{11\to 12}=\rho_{11}^2 M_{11\to 11} M_{11\to 12}^*, \qquad 4 m_1^2< s < (m_1+m_2)^2 \la{extUnit1112}
\eeq

Here, since we are focusing on the top curve (which is crossing equivalent to any of the other two)  we can think of $M$ as a single function of $s$ with 
\begin{eqnarray}
&&t(s)=\frac{1}{2} \left(3
   m_1^2+m_2^2-s-\sqrt{\frac{\left(s-4 m_1^2\right) \left(-2 m_2^2
   \left(m_1^2+s\right)+\left(s-m_1^2\right){}^2+m_2^4\right)}{s}}\right)\label{t11to12}\\
&&u(s)=\frac{1}{2} \left(3
   m_1^2+m_2^2-s+\sqrt{\frac{\left(s-4 m_1^2\right) \left(-2 m_2^2
   \left(m_1^2+s\right)+\left(s-m_1^2\right){}^2+m_2^4\right)}{s}}\right)\label{u11to12}
\end{eqnarray}
As a check, note that as $m_2 \to m_1$ we find $u \to 0$ and $t\to 4m_1^2-s$ as expected for two dimensional elastic scattering of particles of equal mass. 

The extended unitarity relation (\ref{extUnit1112}) is of course part of a coupled system of equations when we consider all components at once. They can all be nicely packed into matrix form by defining 
\beq
\mathbb{U}\equiv 2\IM \mathbb{M}-\mathbb{M}^\dagger \bbr\, \mathbb{M}  \,,
\label{unitarityfullsystem}
\eeq
where
\beq
\!\!\!\! \mathbb{M}\equiv \begin{pmatrix}
M_{11\to 11} & M_{11\to 12} \\
M_{11\to 12} & M_{12\to 12}
\end{pmatrix}, \,\,\,\,\,
\bbr \equiv \begin{pmatrix}
\rho_{11}^2=\frac{\theta\left(s-4m_1^2\right)}{2\sqrt{s-4m_1^2}\sqrt{s}} & 0 \\
0 & \rho_{12}^2=\frac{\theta\left(s-(m_1 + m_2)^2\right)}{2\sqrt{s-(m_1 + m_2)^2}\sqrt{s-(m_1 - m_2)^2}} 
\end{pmatrix} \la{rhoMatrix}
\eeq
Then extended unitarity is the statement that $\mathbb{U}=\mathbf{0}$ for $s\in [4m_1^2,(m_1+m_2)^2]$. Above $s=(m_1+m_2)^2$ we are at physical energies and the extended unitarity relation is replaced by regular unitarity which is now nothing but the statement that $\mathbb{U}$ is a positive semi-definite matrix $\mathbb{U} \succeq 0$ for $s>(m_1+m_2)^2$.\footnote{Strictly speaking we can impose $\mathbb{U}=\mathbf{0}$ for a while longer in the unitarity region, more precisely until the energy where we can  produce two particles of type $2$ or three particles of type $1$. In practice, bounds we will find will saturate unitarity so this will be automatic. Because of this, in all implementations, we will actually impose $\mathbb{U} \succeq 0$ even in the extended unitarity region, that is for any $s>4m_1^2$. This is very convenient as it renders the problem convex. }

\begin{figure}[t]
	\centering 
	\includegraphics[scale=0.5]{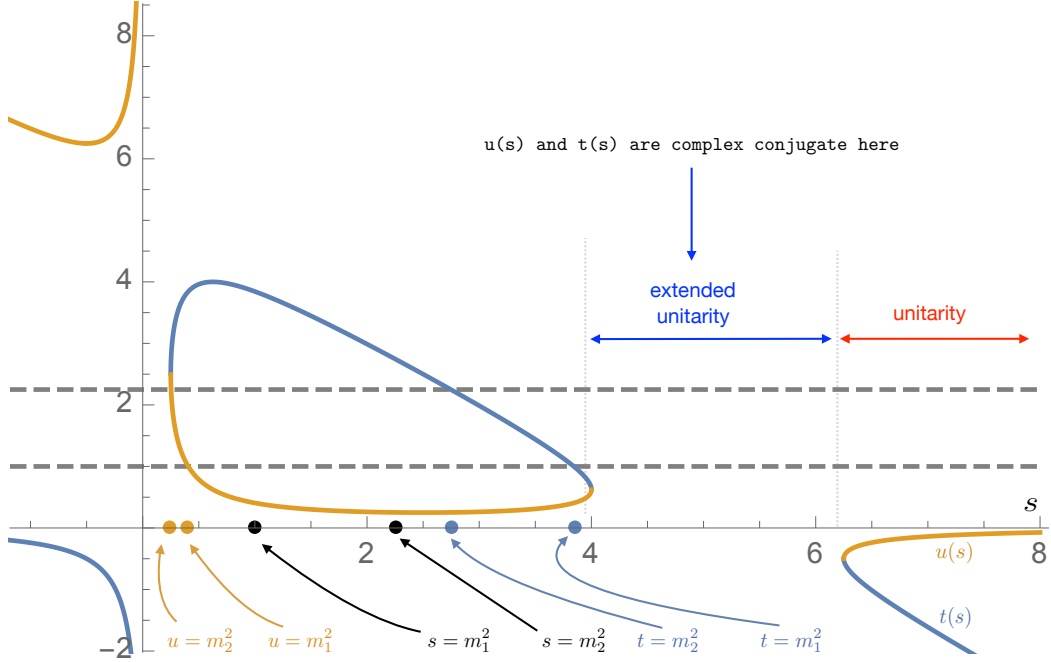}
	\vspace{-1.5cm}
	\caption{$t(s)$ (blue) and $u(s)$ (yellow) for $11\to 12$ scattering and $m_2=\tfrac{3}{2} m_1$. $u(s)$ and $t(s)$ are two branches of the same analytic function. In the extended unitarity region they are complex. As a function of $s$, all poles are located before the extended unitarity region. The grey horizontal dashed lines are equal to $m_1^2$ and $m_2^2$ and fix the position of the $t$-- and $u$-- channel poles.}
	\label{poles1112}
\end{figure}

%Similarly, for the $12\to 12$ component we would have 
%\beq
%2\IM M_{12\to12} = \rho_{11}^2 |M_{11\to 12}|^2 \,, \qquad4m_1^2<s<(m_1+m_2)^2 \,.
%\eeq

\begin{figure}[t]
	\centering 
	\includegraphics[scale=0.5]{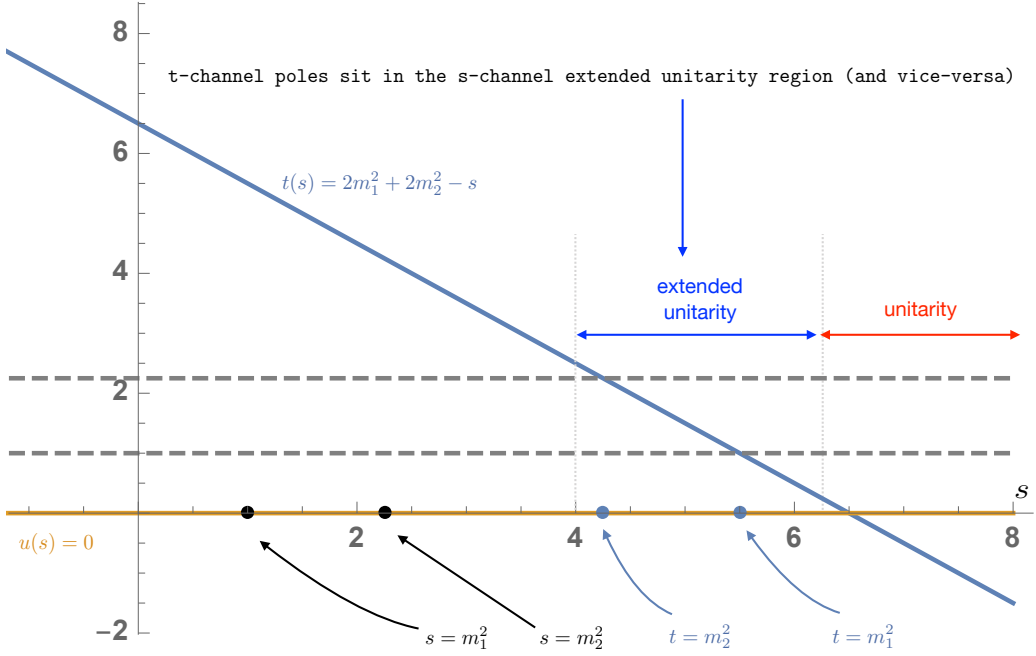}
	\vspace{-1.5cm}
	\caption{$t(s)$ (blue) and $u(s)=0$ (yellow) for $12\to 12$ forward scattering and $m_2=\tfrac{3}{2} m_1$. In the $s$-channel extended unitarity sit $t$-channel poles (and vice-versa). The $s$-channel poles lie before the s-channel extended unitarity region. As in the previous figure, the grey horizontal dashed lines are equal to $m_1^2$ and $m_2^2$ determine the position of $t$-channel poles. }
	\label{poles1212}
\end{figure}

%Above $s=(m_1+m_2)^2$ we are at physical energies and the extended unitarity relation becomes regular unitarity which is most convenient cast in matrix form as \red{eq (\ref{unitarityfullsystem}) is both unitarity AND extended unitarity. it holds for any  $s>4m_1^2$, since the theta functions in (\ref{rhoMatrix}) properly drop the heavy intermediate states for the extended unitarity region. It is a bit mislead what is said in this paragraph, therefore.} 
%\beq
%\mathbb{U}=2\IM \mathbb{M}-\mathbb{M}^\dagger \bbr\, \mathbb{M} \succeq 0\,, \qquad s>(m_1+m_2)^2 \,,
%\label{unitarityfullsystem}
%\eeq
%where
%\beq
%\mathbb{M}=\begin{pmatrix}
%M_{11\to 11} & M_{11\to 12} \\
%M_{11\to 12} & M_{12\to 12}
%\end{pmatrix}, \,\,\,\,\,
%\bbr=\begin{pmatrix}
%\rho_{11}^2=\frac{\theta\left(s-4m_1^2\right)}{2\sqrt{s-4m_1^2}\sqrt{s}} & 0 \\
%0 & \rho_{12}^2=\frac{\theta\left(s-(m_1 + m_2)^2\right)}{2\sqrt{s-(m_1 + m_2)^2}\sqrt{s-(m_1 - m_2)^2}} 
%\end{pmatrix} \la{rhoMatrix}
%\eeq
%
Finally we have poles. These correspond to the single particle exchanges when $s$ or $t$ or $u$ are equal to either $m_1$ or $m_2$.  The poles show up in the (rounded) triangle region in the Mandelstam triangle picture \ref{triangle} in the $11\to 12$ process as depicted in figure \ref{poles1112}. For $12\to 12$, we have $u=0$ and the two $t$-channel poles lie in the extended unitarity region. Note here the important difference between unitarity and extended unitarity. In the unitarity region the amplitudes describe physical probability amplitudes, are bounded and can thus never have poles. In the extended unitarity region they can in principle. And here they do as we see in the figure.  

All in all, we can summarize the analytic structure of our amplitudes with their cuts and poles by dispersion relations as usual. 
These can be conveniently packaged into a simple matrix statement $\mathbb{A}=\mathbf{0}$ with
\beq
\mathbb{A}\equiv \begin{pmatrix}
\mathcal{A}_{11\to11} & \mathcal{A}_{11\to 12} \\
\mathcal{A}_{11\to 12} & \mathcal{A}_{12\to12}
\end{pmatrix} \, \la{AA}
\eeq
and 
\begin{align}
\mathcal{A}_{11\to11}(s)\equiv&M_{11\to11}(s)-M_{11\to11}^\infty+{\color{red}g_{111}^2}\left(\frac{1}{s{-}m_1^2}+\frac{1}{t(s){-}m_1^2}\right)+{\color{blue}g_{112}^2}\left(\frac{1}{s{-}m_2^2}+\frac{1}{t(s){-}m_2^2}\right)\nonumber\\
&- \frac{1}{\pi}\int_{4m_1^2}^\infty \IM M_{11\to11}(z)\left(\frac{1}{z-s}+\frac{1}{z-t(s)}\right)dz\,,\label{m11to11disp} \\
\mathcal{A}_{11\to12}(s)\equiv&\,M_{11\to 12}(s)-M_{11\to12}^\infty+{\color{red}g_{111}}{\color{blue}g_{112}}\left(\frac{1}{s-m_1^2}+\frac{1}{t(s)-m_1^2}+\frac{1}{u(s)-m_1^2}\right)\nonumber\\
&-\frac{1}{\pi}\int_{4m_1^2}^\infty \IM M_{11\to12}(z)\left(\frac{1}{z-s}+\frac{1}{z-t(s)}+\frac{1}{z-u(s)}\right)dz\,,\label{m11to12disp}
\end{align}
\begin{align}
\mathcal{A}_{12\to12}(s)\equiv&\,M_{12\to12}(s)-M_{12\to12}^\infty+{\color{blue}g_{112}^2}\left(\frac{1}{s-m_1^2}+\frac{1}{t(s)-m_1^2}\right)\nonumber\\
&-\frac{1}{\pi}\int_{4m_1^2}^\infty \IM M_{12\to12}(z)\left(\frac{1}{z-s}+\frac{1}{z-t(s)}\right)dz\,.
\label{m12to12disp}
\end{align}
We hope there will be no confusing created by the fact that $t(s)$ signifies different things depending in which equation we are since crossing is implemented differently for different components. In~(\ref{m11to11disp}) is it $t(s)=4m_1^2-s$; in (\ref{m11to12disp}) it is given by (\ref{t11to12}); and in (\ref{m12to12disp}) it is given by~$t(s)=2m_1^2+2m_2^2-s$. In what follows, it should always be clear from the context which~$t(s)$ we are talking about.

\subsection{Multiple Component Dual Problem} \la{mDual}

The formulation of the dual problem for the multiple component scenario can be derived following the steps outlined in Sec.~\ref{sec2}.
There are, however, two practical obstacles: one is the complicated analytic structure of the $11\to12$ component, the other is the presence of the \emph{extended} unitarity region. 
In this section we shall solve both problems if we want to arrive at an elegant and efficient dual numerical setup.

As always, we start from the primal radial problem
\begin{mdframed}[frametitle={Primal Radial Problem for Multiple Component},frametitlealignment=\centering,backgroundcolor=blue!6, leftmargin=0cm, rightmargin=0cm, topline=false,bottomline=false, leftline=false, rightline=false] 
\vspace{-0.4cm}
\begin{align} &\underset{\text{in } {R^2,\mathbb{M}}}{\text{maximize}} &&  R^2\nonumber\\
& \text{constr. by}  && 
0=c_1 \equiv \text{Res}_{m_1^2}(M_{11\to11})- R^2\sin^2\beta \,,   \nonumber\\
& && 0=c_2\equiv \text{Res}_{m_2^2}(M_{11\to11})-R^2\cos^2\beta \,,\nonumber \\
& &&  0=c_3 \equiv \text{Res}_{m_1^2}(M_{11\to12})-R^2\sin\beta\cos\beta\,, \nonumber\\ 
& &&  0=c_4 \equiv \text{Res}_{m_1^2}(M_{12\to12})-R^2\cos^2\beta \,,\nonumber\\
& s > 4m_1^2  && \mathbb{A}=0 \qquad \text{where $\mathbb{A}$ is given in (\ref{AA})} \nonumber \,,  \\
&  s > 4m_1^2&& \mathbb{U} \succeq 0\qquad \text{where $\mathbb{U}$ is given in (\ref{unitarityfullsystem})}\,.
 \label{primal bootstrap 11to12}
\end{align}
\end{mdframed}
%\begin{mdframed}[backgroundcolor=red!6, leftmargin=0cm, rightmargin=0cm, topline=false,bottomline=false, leftline=false, rightline=false] 
%\begin{align} &\underset{\text{in } {R^2,\mathbb{M}}}{\text{maximize}} &&  R^2\nonumber\\
%& \text{constr. by}  && \text{Res}_{m_1^2}(M_{11\to11})=R^2\sin^2\beta, \quad \text{Res}_{m_2^2}(M_{11\to11})=R^2\cos^2\beta \nonumber\\
%& &&  \text{Res}_{m_1^2}(M_{11\to12})=R^2\sin\beta\cos\beta, \quad \text{Res}_{m_1^2}(M_{12\to12})=R^2\cos^2\beta\nonumber\\
%& s > 4m_1^2  && \mathbb{A}=0 \qquad \text{where $\mathbb{A}$ is given in (\ref{AA})} \nonumber\\
%& 4m_1^2 < s < (m_1{+}m_2)^2&& \mathbb{U}^{\text{ext}}(s) \succeq 0,\nonumber\\
%& s> (m_1{+}m_2)^2&& \mathbb{U}(s)\succeq 0.
% \label{primal bootstrap 11to12}
%\end{align}
%\end{mdframed}
If not for the $c_i=0$ equality constraints related to the radial problem, this setup would fit~(\ref{primal2}). 
Note also that the last constraint incorporate automatically unitarity and extended unitarity. Sometimes it is convenient to analyze it separately in the extended and regular unitarity regions corresponding to $s$ bigger/smaller than $(m_1+m_2)^2$ respectively. 

We start our path towards the dual problem with the usual Lagrangian starting point 
%with $\mathbb{U}$ given in eq.~\eqref{unitarityfullsystem} and $\mathbb{U}^{\text{ext}}$ is simply related to  $\mathbb{U}$ setting $\rho_{12}\to 0$.
%The matrix $\mathbb{A}$ packages the dispersion relations~\eqref{m11to11disp},~\eqref{m11to12disp} and~\eqref{m12to12disp}.
%\red{I would not distinguish U for extendend unitarity and unitarity. They are the same constraint, given by (\ref{unitarityfullsystem}-\ref{rhoMatrix}). This will lead to much more compact formulas. And we can put more details under the carpet. If we do that, then the box above is identical to (\ref{primal2}) with the addition of normalization constraints.}
%The ``tour de force'' that will lead us to the multiple components dual formulation starts with the lagrangian
\beq
\mathcal{L}{=}R^2 + \sum_{i=1}^4 c_i \nu _i +
\int_{4m_1^2}^\infty \tr{(\bbw \mathbb{A})}\,ds
%+\int_{4m_1^2}^{(m_1{+}m_2)^2}\tr {(\bbL^{\text{ext}} \mathbb{U}^{\text{ext}})}\,ds
+\int_{4m_1^2}^\infty\tr{(\bbL \mathbb{U})}\,ds,
\label{fullsystlag}
\eeq
%\beq
%\mathcal{L}{=}(\text{rad. constr.})+
%\int_{4m_1^2}^\infty \tr{(\bbw \mathbb{A})}\,ds+
%%+\int_{4m_1^2}^{(m_1{+}m_2)^2}\tr {(\bbL^{\text{ext}} \mathbb{U}^{\text{ext}})}\,ds+
%\int_{4m_1^2}^\infty\tr{(\bbL \mathbb{U})}\,ds,
%\label{fullsystlag}
%\eeq
with $$\bbw =\begin{pmatrix}
w_1 & \tfrac{1}{2}w_2\\
\tfrac{1}{2}w_2 & w_3
\end{pmatrix}$$  
%\beq
%\bbw =\begin{pmatrix}
%w_1 & \tfrac{1}{2}w_2\\
%\tfrac{1}{2}w_2 & w_3
%\end{pmatrix},
%\eeq
and $\bbL$ semi-definite positive. Next we want to identify $\bbw$ as the discontinuities of full analytic functions $\mathbb{W}$ such that the resulting lagrangian becomes manifestly local. This is still possible here but turns out to be more interesting than before because of the richer $11\to 12$ kinematics reviewed in the previous section. The final result is 
\beq\mathbb{W} =\begin{pmatrix}
W_1 & \tfrac{1}{2}W_2\\
\tfrac{1}{2}W_2 & W_3
\end{pmatrix} \label{Wmat}
\eeq
with the dispersive representations of the three \emph{dual scattering functions}
\begin{align}
\label{analyticWcitable}
W_1(s)&=\frac{1}{\pi}\int_{4m_1^2}^\infty dz\, \IM W_1(z)\left(\frac{1}{z-s}-\frac{1}{z-4m_1^2+s}\right),\\
W_2(s)&=\frac{1}{\pi}\int_{4m_1^2}^\infty dz\,\IM W_2(z)\left(\frac{1}{z-s}+\frac{J_t(s)}{z-t(s)}+\frac{J_u(s)}{z-u(s)}\right), \la{W2disp}\\
W_3(s)&=\frac{1}{\pi}\int_{4m_1^2}^\infty dz\, \IM W_3(z)\left(\frac{1}{z-s}-\frac{1}{z-(m_1+m_2)^2+s}\right).
\end{align}
Note that the first and last lines here are pretty much as before: they correspond to anti-crossing symmetric symmetric functionals $W_1$ and $W_3$. The middle line -- with its  Jacobians $J_t=dt/ds$ and $J_u=du/ds$ from (\ref{u11to12},\ref{t11to12}) -- is more interesting and more subtle. We explain its origin in full detail in appendix \ref{W2explanation}. 

%
%
%For the sake of clarity and at the cost of being redundant, we can summarize the dispersive representations of the three \emph{dual scattering functions}
%\begin{align}
%\label{analyticWcitable}
%W_1(s)&=\frac{1}{\pi}\int_{4m_1^2}^\infty dz\, \IM W_1(z)\left(\frac{1}{z-s}-\frac{1}{z-4m_1^2+s}\right),\\
%W_2(s)&=\frac{1}{\pi}\int_{4m_1^2}^\infty dz\,\IM W_2(z)\left(\frac{1}{z-s}+\frac{J_t(s)}{z-t(s)}+\frac{J_u(s)}{z-u(s)}\right),\\
%W_3(s)&=\frac{1}{\pi}\int_{4m_1^2}^\infty dz\, \IM W_3(z)\left(\frac{1}{z-s}-\frac{1}{z-(m_1+m_2)^2+s}\right).
%\end{align}

Then we have the crucial relation required to render the Lagrangian local:
%As promised the analyticity and crossing constraint can be explicitly written as \red{need to define $\mathbb{W}$.}
%\begin{eqnarray}
%&&\int_{4m_1^2}^\infty \tr{(\bbw\,\mathbb{A})}\,ds=\int_{4m_1^2}^\infty \IM \tr{(\mathbb{W}\, \mathbb{M})}\,ds+  \\
%&&\!\!\!\!\!\!\!\!\!\,+\pi\big(\underset{m_1^2}{\text{Res}}(M_{11\to11}) W_1(m_1^2)+\underset{m_2^2}{\text{Res}}(M_{11\to11}) W_1(m_2^2)+\underset{m_1^2}{\text{Res}}(M_{11\to12}) W_3(m_1^2)+\underset{m_1^2}{\text{Res}}(M_{12\to12}) W_2(m_1^2) \big)\nonumber
%\end{eqnarray}
\begin{eqnarray}
\int_{4m_1^2}^\infty \tr{(\bbw\,\mathbb{A})}\,ds &=&\int_{4m_1^2}^\infty \IM \tr{(\mathbb{W}\, \mathbb{M})}\,ds+\pi\big(\underset{m_1^2}{\text{Res}}(M_{11\to11}) W_1(m_1^2)+\underset{m_2^2}{\text{Res}}(M_{11\to11}) W_1(m_2^2) \nonumber\\
&&+\underset{m_1^2}{\text{Res}}(M_{11\to12}) W_3(m_1^2)+\underset{m_1^2}{\text{Res}}(M_{12\to12}) W_2(m_1^2) \big)\nonumber
\end{eqnarray}
Once we plug this relation into our lagrangian (\ref{fullsystlag}) the last line nicely combines with the first two terms there; these terms are the only terms where $R$, $\nu_i$ and the various residues appear.\footnote{{Recall that $R$, the residues and $M(s)$ for $s>4$ are our primal variables, while $\nu_i$ and $W_i(s)$ are our dual variables.}} Maximization with respect to the residues will relate the various functionals $W$ evaluated at the stable particle masses to the lagrange multipliers $\nu_i$ as before while maximization with respect to $R$ will lead to to a linear constraint involving all these functionals which plays the important role of our normalization condition. It reads: 
%
%add \red{we are not adding anything, just redefining variables.} this term to the lagrangian, similarly to the single component case, we can solve the radial constraints for the lagrange multipliers $\nu_i$ and setting \red{better to say: `` maximising with respect to the primal variable $R$ we obtain a bounded dual functional provided...", since this is what we said in this step both in section 2 and single comp. horn., see for example  text around (\ref{norm}). Also, here as well $\nu$ are suddently becoming $W(m^2)$.}
\beq
1+\pi(W_1(m_1)^2 \sin^2\beta+W_1(m_2^2)\cos^2\beta+W_2(m_1^2)\sin\beta\cos\beta+W_3(m_1^2)\cos^2\beta)=0 \,.
\label{RadialCondFull}
\eeq
At this point we already got rid of the lagrange multipliers, the radius and the residues; our (partially extremized) Lagrangian is now a functional of the real and imaginary parts of the amplitudes $\mathbb{M}$ above $4m_1^4$ and of the functionals $W_i$ also for $s>4m_1^2$.  Our dual functional $d$ is therefore the maximization over the amplitudes $\mathbb{M}$ of 
%lagrangian~\eqref{fullsystlag} greatly simplifies \red{I would not call this object a lagrangian since we already maximized over the primal variable R. it is an intermediate object between $\mathcal{L}$  and $d$. I would antecipate the next paragraph saying "to obtain the dual functional d we should now maximise over the amplitude $M$" and follow with $d(w, \lambda) = \text{max}_M\text{ rhs of eqs below}$}:
\beq
d( \mathbb{W},\bbL)= \sup_{\mathbb{M}} \int_{4m_1^2}^\infty  ds \Big( \tr\!(\IM \mathbb{W}\, \mathbb{M})+\tr{(\bbL \mathbb{U}(\mathbb{M}))}\Big)
\label{LindaLagrangia}
\eeq
Since we are dealing with small $2\times 2$ matrices we found it convenient to go to components at this point and also to separate the last integral into its extended and regular unitarity contributions separately. 

For example, using 
\beq
\bbL=\begin{pmatrix}
\lambda_1 & \tfrac{1}{2} \lambda_2\\
\tfrac{1}{2}\lambda_2^* & \lambda_3
\end{pmatrix}\succeq \mathbf{0}, \label{Lmat} 
\eeq
and evaluating the equations of motion for $\RE M_{12\to12}$ and $\IM M_{12\to12}$ in the extended unitarity region we get
\beq
\RE W_3+2\lambda_3=0,\qquad \IM W_3=0.\nonumber
\eeq
These two equations constrain the dual scattering function associated to the $12\to 12$ to have a discontinuity starting at $(m_1+m_2)^2$. Moreover, the semidefinite-positiveness condition on $\bbL$ 
implies\footnote{Second order variations show that the full positive semidefiniteness of $\bbL$ is required for the critical $\mathbb{M}_c$ to be a maximum.} that 
\beq
\lambda_3(s)\geq 0 \qquad \implies \qquad \RE W_3(s)\leq0, \qquad \text{for } 4m_1^2<s<(m_1+m_2)^2.\nonumber
\eeq
We can solve all equations for all amplitude components in both the regular and extended unitarity region for the simple reason that $\mathbb{U}$ is quadratic in $\mathbb{M}$. In this way we get 
$d(\mathbb{W},\bbL)$ which we should now minimize. Its explicit expression is in appendix \ref{dDual}.

We can now minimize first over positive semi-definite $\bbL$ to obtain our final dual functional~$D(\mathbb{W})$. This step is quite non-trivial but leads to a very compact final result:
\begin{mdframed}[frametitle={Dual Radial Problem for Multiple Component},frametitlealignment=\centering,backgroundcolor=red!6, leftmargin=0cm, rightmargin=0cm, topline=false,bottomline=false, leftline=false, rightline=false] 
\vspace{-0.4cm}
\begin{align} &\underset{\text{in } {\mathbb{W}}}{\text{minimize}} \qquad  D^{\text{ext}}(\mathbb{W})+D^{\text{phys}}(\mathbb{W}) \label{dual bootstrap 12system}\\
&D^{\text{ext}}(\mathbb{W})=-\int_{4m_1^2}^{(m_1+m_2)^2}ds\, \frac{|W_2|^2-4 \RE W_1 W_3+|W_2^2-4 W_1 W_3|}{4 \rho_{11}^2 W_3}\nonumber\\
&D^{\text{phys}}(\mathbb{W})=\int_{(m_1+m_2)^2}^\infty ds\,\left(\frac{\RE W_1}{\rho_{11}^2}+\frac{\RE W_3}{\rho_{12}^2}+\sqrt{\frac{|W_1|^2}{\rho_{11}^4}+\frac{|W_3|^2}{\rho_{12}^4}+\frac{|W_2|^2{+}|W_2^2{-}4 W_1 W_3|}{2\rho_{11}^2\rho_{12}^2}} \right) \nonumber\\
%\nonumber\\
& \text{const. by} \quad \RE W_3 \leq 0,\quad \IM W_3=0 \quad\text{for}\quad 4m_1^2\leq s \leq (m_1+m_2)^2 \nonumber\\
&\text{and by}\qquad\, 1+\pi(W_1(m_1)^2 \sin^2\beta{+}W_1(m_2^2)\cos^2\beta{+}W_2(m_1^2)\sin\beta\cos\beta{+}W_3(m_1^2)\cos^2\beta)=0 \nonumber
%\text{Equation }\eqref{RadialCondFull}
%\label{FinalDualFull}
\end{align}
\end{mdframed}

Here, the two contributions $D^{\text{phys}}(\mathbb{W})$ and $D^{\text{ext}}(\mathbb{W})$ correspond to the contributions of regular and extended unitarity. 
%Indeed, we can nicely check that the integrals of the former reduces to that of the later for $\rho_{12} \to 0$ corresponding to eliminating the $12$ intermediate states and keeping $11$ only in the extended unitarity region. 
The last condition is the normalization condition (\ref{RadialCondFull}) and the next-to-last line with the linear inequality constraint is in the end the only remnant of the positive semi-definiteness of the lagrange multiplier matrix $\bbL$. All these constraints can actually be trivialized as we explain in the next section. This will lead to a unconstrained (albeit non-linear) dual minimization problem which we will then solve numerically. 

%The above formulation seem to betray the promise of an unconstrained dual minimization problem. In the next section we shall see how we can trivialize the dual constraints by a suitable choice of the numerical ansatz.

\subsection{Numerical Results}

Now we perform both a primal and a dual numerical exploration to check the correctness of problem~\eqref{dual bootstrap 12system}. 

It what follows we will propose ansatze to parametrize families of dual functionals $W_j$'s. The cleverer the ansatze, the best will the bounds be and the fastest they will converge of course. Clever or not, it is of course important to stress that any ansatze for $W_i$ leads to a totally rigorous exclusion bound. 

The $11\to 11$ dual ansatz is the same used to produce the rigorous dual bounds in figure~\ref{figHorn}
\beq
W_1(s)=\frac{1}{s(4m_1^2-s)} \sum_{n=1}^{N_{\text{max}}}a_n\left(\rho(s)^n-\rho(t)^n\right),
\label{W1ansatz}
\eeq
where $t=4m_1^2-s$, $a_n$ are free variables and $\rho(s)$ is the usual $\rho$-variable foliation introduced in~\cite{Paper3} -- see eq.~\eqref{rhovariabledef} with $m=m_1$.
%\footnote{{\color{blue}We choose always $s_0=2m_1^2$: we have checked other choices of $s_0$ do not significantly improve the numerical results. }}
%\beq
%\rho_2(s)=\frac{\sqrt{4-s_0}-\sqrt{4-s}}{\sqrt{4-s_0}+\sqrt{4-s}}.
%\eeq
This ansatz has the right branch-point discontinuities and it is manifestly anti-crossing symmetric. At infinity it decays as $W_1\sim s^{-5/2}$; in fact, this behavior ensures that the dual objective in~\eqref{dual bootstrap 11to11} is integrable. 
The poles at $s=4m_1^2$ and $s=0$ are not necessary to obtain optimal bounds, but in practice they speed up the numerical convergence.\footnote{We have numerical evidence to believe they are the right singularities the optimal dual scattering function should have. However, it is worth noticing they do not spoil integrability at threshold. We can look at eq.~\eqref{eq13}: the $\int_4^\infty \IM (M(s) W(s)) \,ds$ is integrable if $W(s)\sim 1/(s-4)$ close to threshold because the amplitude vanishes as $M(s)\sim \sqrt{s-4}$. }

For the $11\to 12$ dual ansatz we use
\begin{align}
W_2(s)&=\frac{1}{\sqrt{4m_1^2-s}\sqrt{4m_1^2-t}\sqrt{4m_1^2-u}}\sum_{n=1}^{P_{\text{max}}}b_n \left(\rho(s)^n+J_t(s)\rho(t)^n+J_u(s)\rho(u)^n\right),
\label{W2ansatz}
\end{align}
where $t$ and $u$ are respectively given in~\eqref{t11to12} and~\eqref{u11to12}. Recall also that $J_t=dt/ds$ and $J_u=du/ds$.
At infinity $W_2\sim s^{-3/2}$, therefore the dual objective function~\eqref{dual bootstrap 12system} wouldn't be integrable at infinity. However, it is sufficient to fix two of the $b_n$'s free variables to ensure the $W_2\sim s^{-5/2}$ decay. 
Notice that eq.~\eqref{W2ansatz} has branch point singularities at $s=t=u=4m_1^2$ where the extended unitarity discontinuity in the physical amplitude start.
At the physical threshold $s=(m_1+m_2)^2$ in principle we could add additional singularities such as a pole (similarly to~\eqref{W1ansatz}), however it turns out that numerically
it makes no difference.

\begin{figure}[t]
	\centering 
	\includegraphics[width=\linewidth]{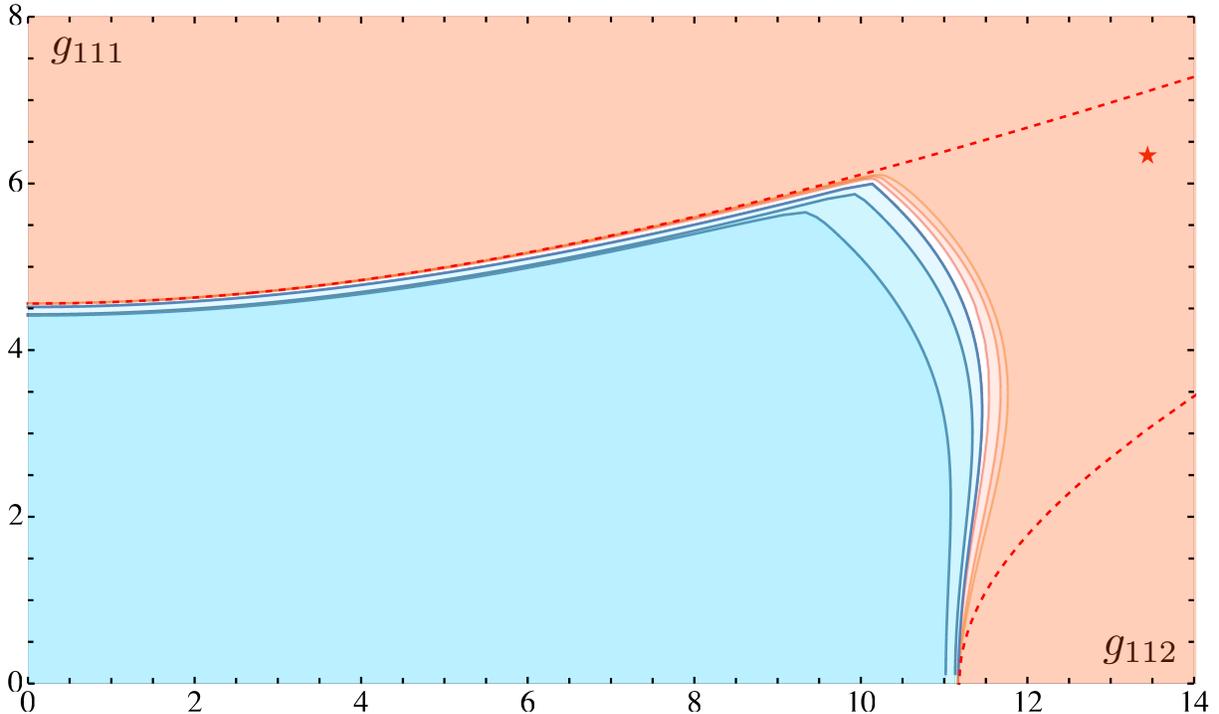}
	\vspace{-3cm}
	\caption{Dual (red) and Primal (blue) excluded/allowed regions once the full system of amplitudes is included.\protect\footnotemark  ~The multi-component improved boundary is now rigorously trapped between the primal and dual bounds. 
	The red dashed line is the previous single component boundary. As we now impose the full system constraints the bound improves dramatically excluding most of the horn like figure. The red star point, for instance, was allowed (feasible) before from the primal problem perspective (it was blue in figure \ref{figHorn}) and is now excluded. Once again, we restrict the plot to the first quadrant due to $g \leftrightarrow -g$ symmetry.}
		\label{figDualFull}
\end{figure}

It is convenient to design the $12\to 12$ dual ansatz such that it automatically satisfy the constraints $\IM W_3=0$ and $\RE W_3<0$ in the extended unitarity region so that our optimization is unconstrained.
The former is easily achieved using a $\rho$-foliation with cut starting at $s=(m_1+m_2)^2$ such as 
\footnotetext{The dual curves, from outer to inner corresponds $(N_\text{max},P_\text{max},Q_\text{max})$ equal to $(8,8,8)$, $(10,10,10)$ and $(10,20,20)$; the primal curves from inner to outer correspond to 136, 271 and 1111 degrees of freedom in the primal ansatz for the amplitude matrix. We used splines analogous to those used in \cite{Paper4}.}
\beq
\tilde\rho(s)=\frac{\sqrt{(m_1+m_2)^2-2m_1^2}-\sqrt{(m_1+m_2)^2-s}}{\sqrt{(m_1+m_2)^2-2m_1^2}+\sqrt{(m_1+m_2)^2-s}}.\nonumber
\eeq
The latter is more subtle: we could always impose linear constraints such as $\RE W_3(s)\equiv W_3(s)\leq 0$ on some grid of points in the $4m_1^2<s<(m_1+m_2)^2$ segment in our dual minimization problem, but this would make \texttt{Mathematica}'s basic \texttt{FindMinimum} slow and nearly unusable.
Instead, we opt to write the ansatz
\beq
W_3(s)=(\tilde\rho(t)-\tilde\rho(s))\left(\frac{1}{\sqrt{(m_1+m_2)^2-s}}+(s{\leftrightarrow} t) \right)\left(\sum_{n=0}^{Q_{\text{max}}}c_n( \tilde\rho(s)^n+\tilde\rho(t)^n) \right)^2\nonumber
\eeq
where $t=2m_1^2+2m_2^2-s$.
It is easy to check that $W_3$ has actually definite sign in a larger region than extended unitarity: $W_3>0$ in $t((m_1+m_2)^2)=(m_1-m_2)^2<s<m_1^2+m_2^2$ 
and $W_3<0$ in $m_1^2+m_2^2<s<(m_1+m_2)^2$ which of course include the extended unitarity region. 
This may sound too restrictive, however this is one of the advantages of the dual formulation:
as long as the dual scattering functions satisfy the dual constraints, the bounds obtained are rigorous.
Of course, a legitimate question is whether our ansatz is able to attain the optimal value of the dual problem.
It turns out that for the case we are studying this ansatz is also approximately optimal numerically. 

%We don't need to worry about being too restrictive since the bound we get are nonetheless rigorous! It turns out that for the case we are studying this ansatz is also \red{approximately (unless you can argue more)} optimal \red{numerically}. 

Now we have all the ingredients to just code the objective in~\eqref{dual bootstrap 12system} and minimize it unconstrained.
The result for the $\{g_{111}, g_{112}\}$ space is shown in figure~\ref{figDualFull} (red shaded regions). In the same figure, the blue shaded areas are determined running the primal problem eq.~\eqref{primal bootstrap 11to12} -- see~\cite{Paper4} for details about primal multiple component numerics. The red dashed line marks the single component analytic bound. The white space in between the primal and dual 
areas is the uncertainty we have in the definition of the boundary for the full coupled system. Clearly the optimal bound is almost completely trapped!

\section{Discussion}

Icarus said that \textit{all limits are self-imposed}. That is not totally true. Unitarity, crossing symmetry and analyticity clearly also impose very important bounds.

In this paper we initiated a general dual bootstrap program and applied it on the next-to-simplest S-matrix bootstrap scenario: Two dimensional amplitudes with more than one particle type and more than one mass.\footnote{The simplest example was kicked off in \cite{Monolith} for a single particle species transforming in some global symmetry group.}

One main goal of this paper was to set up the theory behind this physical problem and connect it with the standard language of dual and primal maximization problems as optimization problems. Indeed, a great deal of section \ref{sec2} can bet transported from (\textit{the continuum limit of}) business and finance department reviews of optimization problems (beautiful examples are \cite{reviews}), or math books \cite{mathbook,mathbook2,mathbook3}. 

In the S-matrix bootstrap studied here the primal problem is linear but constrained; the dual problem is non-linear but unconstrained.\footnote{The unconstrained nature of the dual problem is an extremely powerful and fortunate property which was not a priori guaranteed. It is the nature of the S-matrix Bootstrap problems considered up to now that allowed us to trivialize all dual constraints encountered thus far.} For the primal problem, we used the powerful \texttt{SDPB} code to perform the optimizations. For the dual problem we used \texttt{Mathematica}'s basic \texttt{FindMinimum}.\footnote{\texttt{FindMinimum} is sometimes an art. It is not uncommon to ask for a minimization, give \texttt{Mathematica} a viable starting point and obtain a final result bigger than the starting value. Go figure. Of course, it is a price to pay when having a one size fits all algorithm. See also next footnote.} 
Even so, the dual problem is orders of magnitude faster right now.\footnote{The dual curves in figure \ref{figDualFull} contain thousands of points and take about a day to generate in a regular laptop. The primal curves take a few days in a cluster. One reason why we did not use the cluster for the dual problem is that we found it useful to hotstart \texttt{FindMinimum} by starting the minimization search at a given point using the final result of the neighbouring point.} It would be very interesting to look for more tailor made algorithms for our kind of minimizations to speed the dual even more. 

Of course, the main advantage of having a dual problem is not speed but the fact that the bounds whence generated are completely rigorous. What is once excluded can never be included back. This is in contradistinction with the primal formulation where more constraints will often rule out a previously feasible solution.
 In practice the best is to use both dual and primal problems at once. When they almost touch each other -- meaning the so called duality gap is closing -- we know we are reaching the very optimal bounds!

Having developed the theory and a very fast dual problem, we look forward to putting it to use in several interesting physical applications. 

\begin{figure}[t]
	\centering 
	\includegraphics[width=\linewidth]{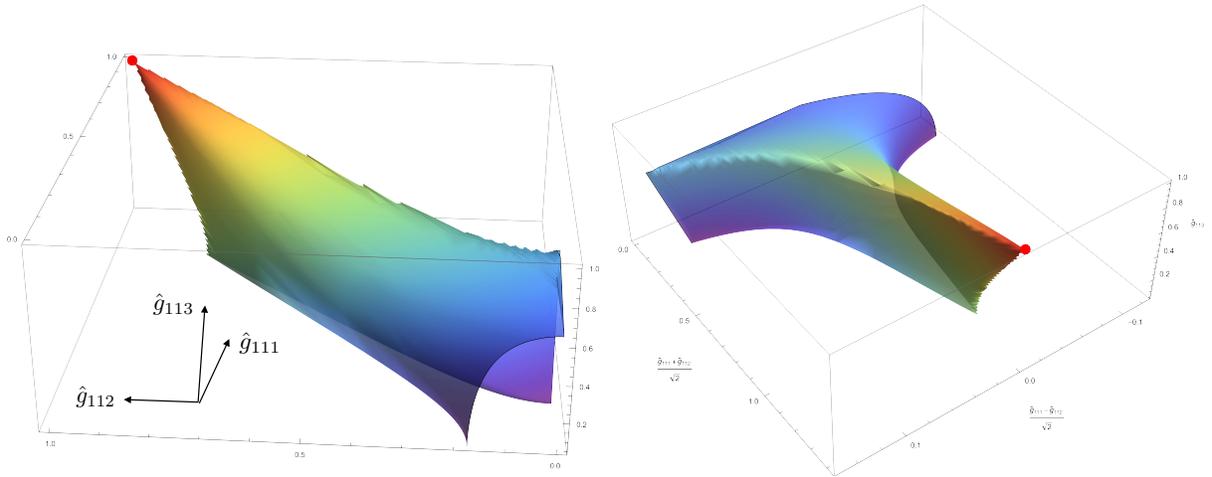}
	\vspace{-6cm}
	\caption{\textbf{Left:} Maximum couplings $g_{11j}$ for a theory with the masses of the Ising field theory deformed by magnetic field depicted by normalizing those by the Ising couplings, $\hat g_{11j}\equiv g_{11j}/g_{11j}^\texttt{Ising}$ at this $E_8$ point. \textbf{Right:} The plot on the right is obtained by a simple rotation of the first by 45 degrees which magnify some of the nice features of the plot. (These plots were generated using the dual method developed in this paper with $N=20$; it might be possible to derive this shape analytically. We did it for one of the faces but did not pursue this further.) 
The Ising field theory, the red dot, lies beautifully at the very tip of these horn shaped single component plots. 
	}
	\label{cusp3D}
\end{figure}

One goal would be to bootstrap the Ising model field theory with both thermal and magnetic deformations turned on. Let us recall why we think this is promising. The Ising field theory with pure magnetic deformation~\cite{ZamoE8} is at the boundary of the single amplitude bound \cite{Paper2}, see figure 12 there. What is more, it is precisely at the top of a sharp horn like 3D bound in the coupling space as depicted in figure \ref{cusp3D}.\footnote{In \cite{Paper2} only the maximum $g_{111}$ coupling was plotted so it was not possible to see this cusp so sharply.} Something we clearly learned in this paper is how multiple amplitudes can truncate such horns; compare figures \ref{figHorn} and \ref{figDualFull}. At the magnetic Ising point this dramatic truncation can not happen. This theory exists after all, we can not rule it out. What happens is that the very special values of the masses of the stable particles of this theory allow for fine tuned cancelations in $11\to 12$ and other amplitudes such that they completely vanish and thus do not affect the single component bound which produces the horn. In other words, the purely magnetic deformation, being precisely integrable, is very special. As soon as we move away from these special masses by turning a thermal deformation, the multiple amplitude bounds are now expected to strongly affect the single component analysis and this provides a strong improvement over the bounds in \cite{Paper2}. This is not totally trivial to implement because close to the magnetic point, the Ising field theory has three stable particles. Exploring the space of couplings $g_{ijk}$ between these particles is hard because this space is ten dimensional. The trick here is to find a clever lower dimensional section of this multidimensional space, with good optimization targets, which could efficiently isolate the magnetic plus thermal Ising deformation. That is something we are currently exploring.\footnote{Using the form-factor bootstrap~\cite{formfactors} as a further complementary tool to nail down the relevant physical Ising deformation might be very powerful as well. And recent work \cite{barak} provides valuable insight into particle production and analytic properties of the expected S-matrices away from the integrable points.}

Another interesting theory to explore would be the tri-critical Ising model. In the discussion section of \cite{Paper4} an S-matrix bootstrap homework exercise was proposed in relation to this model. With the great speed gains from the dual technology here developed this homework seems very doable. The deformation proposed there concerns a deformation preserving $\mathbb{Z}_2$ symmetry. The dual $\mathbb{Z}_2$ symmetric bootstrap is discussed in appendix \ref{Z2appendix} for the case of equal masses; the uneven masses case should be a straightforward generalization of the analysis of the main text. 

\begin{figure}[t]
	\centering 
\includegraphics[width=\linewidth]{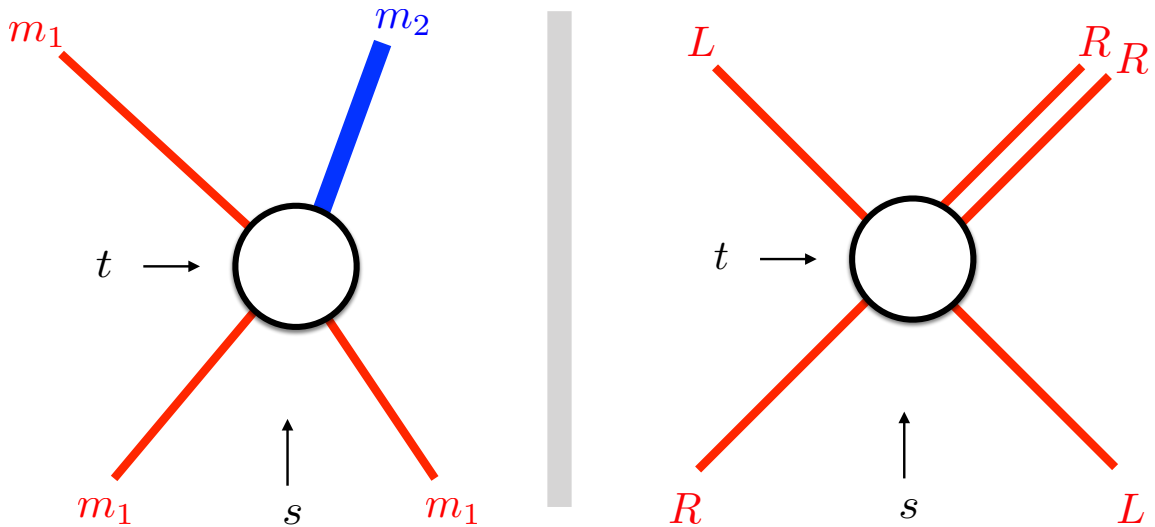}
\vspace{-5cm}
	\caption{The kinematics of the two-to-two process $11\to 12$ is very reminiscent of two-to-three scattering of massless particles as illustrated here. Both processes are fully crossing symmetric. Particle $2$ on the left is analogous to the jet of two right-movers on the right. This two-to-three scattering process should show up in flux tube physics~\cite{Dubovsky1,Zohar,axion} where parity is broken. Extending the flux tube S-matrix boostrap program initiated in \cite{FluxTube} to include such processes would be extremely interesting.
	}
	\label{analogy}
\end{figure}

One step up in the complexity ladder of bootstrap problems are problems whose amplitudes depend on more than a single complex variable. One example is of course higher dimensions where we have both an energy and an angle even in two-to-two scattering processes.\footnote{It is also in higher dimensions where the tension between absence of particle production and crossing symmetry is most striking \cite{aks,amitesasha} which is another point a dual formulation should be very helpful in clarifying.} Another example are higher point amplitudes, even in two dimensions. In fact, in a very roundabout way, we arrived at the class of problems presented in this paper precisely while starting to tackle these multi-particle problems in work in progress with J. Penedones. The point is that the $11\to 12$ amplitudes studied in this paper are in a sense very similar to a sort of $2\to 3$ scattering process of massless particles as illustrated in figure~\ref{analogy}. The jet of the two right movers in the future is like particle $2$. Of course, that jet can have any sub-energy hence we have in that case a continuum of "particles" of type 2, hence the additional complex variable. Nonetheless, this problem seems within reach. We hope to reach it and report on it in the near future. 

 \section*{Acknowledgements}
We would like to specially thank Joao Penedones for enlightening discussions, for collaboration at the initial stages of this project, and for collaboration in several related projects. 
%We would like to thank YYY for several enlightening discussions. 
%
Research at the Perimeter Institute is supported in part by the Government of Canada 
through NSERC and by the Province of Ontario through MRI. This work was additionally 
supported by a grant from the Simons Foundation (PV: \#488661) and FAPESP 
grants 2016/01343-7 and 2017/03303-1.

\appendix

%\section{Connection with Cordova et al formulation}

%\newpage
%\section{Algebra}

\section{Strong Duality}
\label{strong}

Assume $M_*$ solves the primal problem (\ref{primal2}) with optimal coupling $g^2_*$, and note that there are some amplitudes do not saturate unitarity since we could always cook up models with other particles also with mass $m_1$ and $m_2$ so that probability could leak into those \textit{hidden} sectors and manifest itself as non-saturation of unitarity in our truncated Hilbert space.\footnote{Here unitarity refers to both unitarity and extended unitarity.} This means that

                  \begin{mdframed}[frametitle={Inner point property},frametitlealignment=\centering,backgroundcolor=black!10, leftmargin=0cm, rightmargin=0cm, topline=false,
	bottomline=false, leftline=false, rightline=false] 
	\vspace{-0.1cm}
	\begin{equation}
\text{ There exists an $M_i$ such that $ \mathbb{U}(M_i) \succ 0 \text{ and } \mathbb{A}(M_i) = 0$. }\label{innerpoint}
 \end{equation}
	     \end{mdframed}
	    
In this appendix we argue that this implies strong duality \cite{mathbook, mathbook2, mathbook3}. Consider the following auxiliary convex set in the space of real $\mathcal{G}$, symmetric $\mathcal{A}(s)$ and hermitian $ \mathcal{U}(s)$:
\beq
\textbf{Aux} = \{(\mathcal{G}, \mathcal{A}(s), \mathcal{U}(s))\text{ s.t. } \mathcal{G} \leq g^2, \label{aux} \mathcal{A}(s) = \mathbb{A}(M(s)), \mathcal{U}(s) \preceq \mathbb{U}(M(s)) \text{ for some analytic M}\}.\nonumber
\eeq
 The point $\textbf{bp} \equiv \{g^2_*, 0, 0\}$ is at the boundary of $\textbf{Aux}$. Since $\textbf{Aux}$ has an interior point, the supporting hyperplane theorem\footnote{In the infinite dimensional case, this is a consequence of Hahn-Banach's theorem.} guarantees that there exists a hyperplane (i.e. a linear functional on the $(\mathcal{G}, \mathcal{A}(s), \mathcal{U}(s))$ space) containing $\textbf{bp}$ so that the set $\textbf{Aux}$ is to one side of it. In equations, there exists real $\gamma$, symmetric $ \bbw_c(s)$ and hermitian $\bbL_c(s)$, not all simultaneously zero, such that
 \beq
 \gamma\mathcal{G} + \int_{4m_1^2}^\infty ds \text{ Tr} ( \bbw_c \cdot \mathcal{A}(s) + \bbL_c \cdot \mathcal{U}(s) ) \leq \gamma g^2_* \qquad \text{ for all $(\mathcal{G}, \mathcal{A}(s), \mathcal{U}(s))$ in $\textbf{Aux}.$} \label{steq1}
 \eeq
 Note that, due to the definition of $\mathcal{A}$, this is only possible if $\gamma \geq 0$ and $\bbL \succeq 0$.\footnote{To see explicitly, assume $\bbL_c(s)$ isn't positive semidefinite. Then there exists $\vec{x}$ such that $\vec{x}^\dagger \bbL_c \vec{x} < 0$.  In turn, this implies that $ \text{ Tr}( \bbL_c \cdot ( \mathcal{U}  + r \vec{x} \vec{x}^\dagger) ) $ could become arbitrarily negative as we take $r \rightarrow \infty$. Note, however, that $(\mathcal{G}, \mathcal{A}, \mathcal{U}  + r \vec{x} \vec{x}^\dagger)$ is in $\mathcal{A}$ for all $r>0$ provided $(\mathcal{G}, \mathcal{A}, \mathcal{U} )$ is. These two facts together are in contradiction with (\ref{steq1}).}  
 
 Equation (\ref{steq1}) should hold in particular when the inequalities in (\ref{aux}) are saturated, in which case it reduces to
 \beq
 \gamma g^2 + \int_{4m_1^2}^\infty ds \text{ Tr} ( \bbw_c \cdot \mathbb{A} + \bbL_c \cdot \mathbb{U}) \leq \gamma g^2_* \qquad \text{ for all $M$} \label{steq2}.
 \eeq
 Next, we need to argue that $\gamma \neq 0$. First note that, if that were the case, then, after renormalizing $\bbw_c \rightarrow \gamma \bbw_c$ and $ \bbL_c \rightarrow \gamma \bbL_c$, we would conclude that
 \beq
 \mathcal{L}(M, \bbw_c, \bbL_c) =  g^2 + \int_{4m_1^2}^\infty ds \text{ Tr} ( \bbw_c \cdot \mathbb{A} + \bbL_c \cdot \mathbb{U}) \leq  g^2_* \qquad \text{ for all $M$},\nonumber
 \eeq
which, paired  with weak duality, leads to strong duality:  
 \beq
g_*^2 \leq d(\bbw_c,\bbL_c) \equiv  \underset{\mathbb{M}}{\text{sup  }} \mathcal{L}(\mathbb{M}, \bbw_c ,{\bbL_c})  \leq g_*^2 \implies d(\bbw_c,\bbL_c) = g_*^2.\nonumber
 \eeq
 In particular this implies that unless $\bbL_c$ has a zero eigenvalue, $\mathbb{U}=0$, i.e. unitarity is saturated. This is the matrix version counterpart of the argument in \cite{Monolith} for unitarity saturation.
 
 To prove that $\gamma > 0$, assume $\gamma = 0$ and look for a contradiction. Plugging $M_i$ from (\ref{innerpoint}) into (\ref{steq2}) would show that $\bbL_c = 0$. This in turns would lead, using (\ref{steq1}), to 
 \beq
  \int_{4m_1^2}^\infty ds \text{ Tr} ( \bbw_c \cdot \mathcal{A}(s)) \leq 0 \qquad \text{ for all $(\mathcal{G}, \mathcal{A}(s), \mathcal{U}(s))$ in $\textbf{Aux}.$},\nonumber
 \eeq
 which can only be true for a symmetric $\bbw$ if $\bbw_c =0$. But $\gamma,  \bbw_c(s), \bbL_c(s)$ are not all zero by the supporting hyperplane theorem, which shows that $\gamma =0$ is a contradiction. This concludes the argument.

\section{More on dispersion relations}
\subsection{Subtracted dispersions}
\label{analyticstuff}

The construction of the dual problem starts with the dispersive representation of the amplitude, see for instance eq.~\eqref{analandcrossing}.
In order to allow the most general behavior compatible with polynomial boundedness, one introduces subtractions.

Here we show that for the case of $M=M_{11\to 11}$ scattering, our derivation is compatible with one-subtracted dispersions.
Let us start from the identity (we set the units by the mass of the external particle $m=1$)
\beq
M(s)-M(2)=\int_{\mathcal{C}_\varepsilon(s)} \frac{M(z)}{z-s}dz-\int_{\mathcal{C}_\varepsilon(2)} \frac{M(z)}{z-2}dz,
\label{subdispersions}
\eeq
where $\mathcal{C}_\varepsilon(s_0)$ is a circular path around $s_0$ of radius $\varepsilon$ and we can imagine there always exist a path connecting them.
The amplitude $M(s)$, by physical assumptions, can only have poles on the real axis in the segment $0<s<4$.\footnote{We chose $s=2$ as a subtraction point for convenience: the only dangerous situation is when the mass of the bound state is $m_b^2=2$.
However, in that case the $s$ and $t$-channel poles would collide canceling each other, therefore we can avoid this situation and always assume $m_b^2\neq 2$ without loosing generality.} 

Blowing up the contour in eq.~\eqref{subdispersions} we get
\begin{align}
0=\mathcal{A}(s)&=M(s){-}M(2)-\sum_i g^2_i(s-2)\left(\frac{1}{(m_i^2-s)(m_i^2-2)}-\frac{1}{(t(m_i^2)-s)(t(m_i^2)-2)} \right)\nonumber\\
&-\frac{1}{\pi}\int_{4}^\infty \IM M(z)\left( \frac{s-2}{(z-s)(z-2)}+\frac{t(s)-t(2)}{(z-t(s))(z-t(2))} \right) dz\nonumber\\
&=M(s){-}M(2)+\sum_i g_i^2\left(\frac{1}{s-m_i^2}{+}\frac{1}{t(s)-m_i^2}{-}\frac{2}{m_i^2-2}\right)\nonumber\\
&-\frac{1}{\pi}\int_{4}^\infty \IM M(z)\frac{2(s{-}2)^2}{(z{-}s)(z-2)(z{+}s{-}4)} dz.
\label{subdispersions2}
\end{align}
The last line of equation above shows that the imaginary part of the amplitude can grow as $\IM M(z)\sim z$ for large $z$.

We want to integrate~\eqref{subdispersions2} against the Lagrange multiplier $w(s)$. Now, note that a new {primal} variable we have now is the constant $M(2)$ in (\ref{subdispersions2}); when we construct the Lagrangian by integrating (\ref{subdispersions2}) agains $w(s)$, that constant term will be multiplied by the integral of $w(s)$ and thus its equations of motion will lead to $\int_4^\infty w(s)ds=0$ which we assume henceforth. Then, it is easy to show that
%{\color{magenta} Remove: If we choose the multiplier such that $\int_4^\infty w(s)ds=0$, it is easy to show that}
\begin{align}
\int_4^\infty w(s)\mathcal{A}(s)ds&=\int_4^\infty M(s)w(s)+\sum_i g_i^2\int_4^\infty w(s)\left(\frac{1}{s-m_i^2}+\frac{1}{t(s)-m_i^2}\right)ds\nonumber\\
&-\frac{1}{\pi}\int_4^\infty dz\,\IM M(z) \int_4^\infty\, w(s) \frac{2(s{-}2)^2}{(z{-}s)(z-2)(z{+}s{-}4)}.\nonumber
\end{align}
If we decompose the subtracted integration kernel in partial fractions 
\beq
\frac{2(s{-}2)^2}{(z{-}s)(z-2)(z{+}s{-}4)}=\frac{1}{z-s}+\frac{1}{z+s-4}-\frac{2}{z-2}\nonumber
\eeq
the integration against the Lagrange multiplier nicely yields\footnote{In all these manipulations we are assuming that $w(s)$ decays fast enough to justify the integration exchanges.}
\beq
 \int_4^\infty\, w(s) \frac{2(s{-}2)^2}{(z{-}s)(z-2)(z{+}s{-}4)}ds= -\int_4^\infty\, w(s) \left(  \frac{1}{s-z}-\frac{1}{s-t(z)}\right)ds.\nonumber
\eeq
Following the logic outlined in Sec.~\ref{sec2}, we introduce the anti-crossing analytic function, holomorphic in the complex-plane without the 
normal unitarity cuts
\beq
W(z)=\frac{1}{\pi}\int_4^\infty\, \IM W(s) \left(  \frac{1}{s-z}-\frac{1}{s-t(z)}\right)ds,\nonumber
\eeq
such that $\IM W(s)=w(s)$ for $s>4$.
At the end we get the useful identity
\beq
\int_4^\infty w(s)\mathcal{A}(s)ds=\int_4^\infty \IM (W(s) M(s))ds +\pi\sum_i g_i^2 W(m_i^2),\nonumber
\eeq
that we have used, for instance, to get eq.~\eqref{eq13}.

\subsection{The $11\to 12$ functional.} \la{W2explanation}
The analysis leading to the dispersion relation (\ref{W2disp}) arises from the analysis of the term 
 %\red{below, either use $Res(M)$ or $g_111$. We are alternating}
%The radial constraints terms are analogous to the ones introduced in~\eqref{lagrangianhorn}
%\begin{align}
%(\text{rad. constr.})=&R^2+\nu_1(\text{Res}_{m_1^2}(M_{11\to11})-R^2\sin^2\beta)+\nu_2(\text{Res}_{m_2^2}(M_{11\to11})-R^2\cos^2\beta)\nonumber\\
%&+\nu_3(\text{Res}_{m_1^2}(M_{11\to 12})-R^2 \sin\beta \cos\beta)+\nu_4(\text{Res}_{m_1^2}(M_{12\to12})-R^2\cos^2\beta).
%\end{align}
%\red{I would delete everything from here until (\ref{analyticWcitable}) and add a short footnote about anticrossing symmetry for equal masses vs more general structure in 11->12. I didn't need to go over all of this details for single component and I don't see why we would be interested in doing this here.}
%Analogously to eq.~\eqref{disp} we wish to introduce a matrix of analytic functions that trivializes the analyticity and crossing constraints.
%Let's expand the relevant term in eq.~\eqref{fullsystlag}
%\beq
%\int_{4m_1^2}^\infty \tr{(\mathbf{w} \mathbb{A})}\,ds=\int_{4m_1^2}^\infty \mathcal{A}_{11\to11} w_1+ \mathcal{A}_{11\to12} w_2+ \mathcal{A}_{12\to12} w_3\, ds.
%\eeq
%The ``hard'' constraint is given by the term $\mathcal{A}_{11\to12}w_2$ associated to the $11\to 12$ component
\begin{align}
&\int_{4m_1^2}^\infty  \mathcal{A}_{11\to12} w_2= \text{Res}_{m_1^2}(M_{11\to12}) \int_{4m_1^2}^\infty w_2\left(\frac{1}{s-m_1^2}+\frac{1}{t-m_1^2}+\frac{1}{u-m_1^2}\right)ds+\nonumber\\
&+\int_{4m_1^2}^\infty \RE M_{11\to12} w_2\,ds-\frac{1}{\pi}\int_{4m_1^2}^\infty dz \IM M_{11\to12}(z)\int_{4m_1^2}^\infty w_2(s)\left( \frac{1}{z-s}+\frac{1}{z-t(s)}+\frac{1}{z-u(s)}\right).
\label{hardconstraint}
\end{align}
once we use the dispersion relation (\ref{m11to12disp}). 

The second line suggests that we could define an analytic function $W_2$ such that $w_2=\IM W_2$, in particular 
\beq
\RE W_2(z)=-\frac{1}{\pi}\int_{4m_1^2}^\infty \IM W_2(s)\left( \frac{1}{z-s}+\frac{1}{z-t(s)}+\frac{1}{z-u(s)}\right)\,ds.\nonumber
\eeq
It is interesting to notice that while $M_{11\to12}$ is manifestly crossing invariant in $s,t,u$ because the integration kernel is, the crossing properties of $W_2$ are now implicitly defined and we need to invert the relation between $t(s)$, $u(s)$ and $z$. Some simple algebra shows that
\beq
\RE W_2(z)=\frac{1}{\pi}\int_{4m_1^2}^\infty \IM W_2(s)\left( \frac{1}{s-z}+\frac{J_t(z)}{s-t(z)}+\frac{J_u(z)}{s-u(z)}\right)\,ds,
\label{dualcrossing}
\eeq
with $J_t=dt/ds$ and $J_u=du/ds$.
In other words, we can define an analytic function which is \emph{dual crossing} symmetric in the sense that when we cross we pick a jacobian factor.
Notice that this definition is compatible with~\eqref{disp} as for single component $J_t=dt/ds=-1$ and $J_u=0$. The standard anti-crossing case thus follows as a particular case from this general rule.
From eq.~\eqref{dualcrossing} we immediately recover
\beq
\int_{4m_1^2}^\infty \IM W_2(s)\left(\frac{1}{s-m_1^2}+\frac{1}{t(s)-m_1^2}+\frac{1}{u(s)-m_1^2}\right)ds=\pi W_2(m_1^2).\nonumber
\eeq
In practice, we have shown that eq.~\eqref{hardconstraint} can be reduced to 
\beq
\int_{4m_1^2}^\infty  \mathcal{A}_{11\to12} w_2=\int_{4m_1^2}^\infty \IM (W_2 M_{11\to12})\,ds+\text{Res}_{m_1^2}(M_{11\to12})W_2(m_1^2).\nonumber
\eeq
The analysis of the $12\to12$ component follows straightforwardly and is analogous to the $11\to11$ case.

\section{Dual Lagrangian for multiple components}
\label{dDual}
Here we present some of the algebra manipulations pertaining to section \ref{mDual}. In particular, the final expressions in this appendix contain the optimal phase shifts in terms of the critical dual functionals.
Varying the Lagrangian 
%\beq
%\mathcal{L}(\mathbb{M},\mathbb{W},\bbL)=\int_{4m_1^2}^\infty  ds \Big( \tr\!(\IM \mathbb{W}\, \mathbb{M})+\tr{(\bbL \mathbb{U}(\mathbb{M}))}\Big).
%\eeq
\beq
\mathcal{L}(\mathbb{M},\mathbb{W},\bbL)=\int_{4m_1^2}^\infty  ds\, \tr\!\! \Big( \underbrace{\frac{\mathbb{W}\cdot \mathbb{M}- \overline{\mathbb{M}}\cdot \overline{\mathbb{W}}}{2i}}_{\text{Im}(\mathbb{W}\, \mathbb{M})}
%\cdot \mathbb{M}
%\IM \mathbb{W}\, \mathbb{M})+
+\underbrace{\bbL\cdot \big(2\frac{ \mathbb{M}-\overline{ \mathbb{M}}}{2i}- \mathbb{M} \cdot \rho \cdot \overline{\mathbb{M}}\big)}_{\bbL \cdot \mathbb{U}(\mathbb{M}))} \Big). \la{LagD}
\eeq
with respect to $\mathbb{M}$ and its conjugate\footnote{Note that for the symmetric matrices $\mathbb{M}$ and $\mathbb{W}$ in (\ref{rhoMatrix}) and (\ref{Wmat}) hermitian conjugation is the same as conjugation hence the absence of daggers in these expressions.}
\begin{eqnarray}
0 = \int_{4m_1^2}^\infty  ds \,\tr \Big( \delta \mathbb{M} \cdot { \Big[ \frac{\mathbb{W}}{2i} +\frac{\bbL}{i}  - \rho \cdot \overline{\mathbb{M}} \cdot \bbL  \Big] }+ \delta \overline{\mathbb{M}} \cdot { \Big[ \frac{\overline{\mathbb{W}}}{-2i} +\frac{\bbL}{-i}  - \bbL\cdot \mathbb{M}\cdot \rho \Big] } \Big)\nonumber
\end{eqnarray}
Now, since $\delta M$ (and its hermitian conjugate) are expressed in a basis of pauli matrices $ \sigma_0 (= \mathbb{I}), \sigma_1,\sigma_3$ but \textit{not} $\sigma_2 \equiv \sigma$ we can only say that each term in square brackets is zero up to a term proportional to $\sigma$ which will always vanish under the trace, 
\beq
\frac{\overline{\mathbb{W}}}{-2i} +\frac{\bbL}{-i}  - \bbL\cdot \mathbb{M}\cdot \rho = \bf{a}\, \sigma \,. \la{aEq}
\eeq
At this junction we will split the analysis into the extended and regular unitarity region for the simple reason that $\rho$ is invertible only in the regular unitarity region. 

Let us first focus on the regular unitarity region. Dotting (\ref{aEq}) with $\sigma\cdot\rho\cdot \bbL^{-1}$ from the left and taking the trace kills the last two terms on the left hand side and leads to a simple expression for $\textbf{a}$. Next, armed with $\textbf{a}$ we can simply multiply the equation by $\bbL^{-1}$ and $\rho^{-1}$ from the left/right respectively to get $\mathbb{M}$, 
\beq
\mathbb{M}= \frac{i}{2} \bbL^{-1} \cdot \Big( 2 \bbL+\overline{\mathbb{W}}+ \sigma \underbrace{\frac{\tr\!(\sigma\cdot \mathbb{W}\cdot \bbL^{-1} \cdot \rho)}{\tr\!(\bbL^{-1} \cdot \rho)}}_{\textbf{a}} \Big) \cdot \rho^{-1}\nonumber
\eeq
 We could still simplify this expression a bit more noting that $\rho^{-1}=\sigma\cdot \rho\cdot \sigma/\det(\rho)$ and $\bbL^{-1}=\sigma\cdot \bbL\cdot \sigma/\det(\bbL)$ to get rid of the some inverses. Finally, we can plug this expression into the Lagrangian (\ref{LagD}) to obtain a beautiful compact matrix form for the dual objective in the regular unitarity region: 
\beq
d^{\text{regular}}(\mathbb{W},\bbL){=} \int\limits_{4m_1^2}^\infty  ds\, \tr\!\! \Big( \rho^{-1} \cdot \Big( \bbL+\frac{\mathbb{W}}{2}\Big)\cdot \bbL^{-1} \cdot  \Big( \bbL+\frac{\overline{\mathbb{W}}}{2}\Big)  \Big)+\frac{\tr\!\! \big(\rho^{-1}{\cdot} \mathbb{W}{\cdot} \sigma{\cdot} \bbL\big)\tr\!\! \big(\rho^{-1}{\cdot} \overline{\mathbb{W}}{\cdot} \bbL^{-1} {\cdot} \sigma\big)}{4\,\tr\!\! \big(\rho^{-1}\cdot \bbL\big)} \,. \nonumber
\label{dregular}
\eeq
Nicely, note how one can formally reduce it to a single component by replacing $\sigma$ by zero (thus killing the second term), dot products by simple products and matrices by functions; then this Lagrangian would precisely reduce to the single component expression (\ref{dWL}). 

Next we consider the extended unitarity region. An annoying feature is now that $\rho$ is not invertible. On the other hand, the reason why $\rho$ is not invertible is precisely because it becomes full of zeros and hence extremely simple: 
\beq
\rho \to \left(
\begin{array}{cc} 
\rho_{11}^2 & 0 \\
0 & 0 \end{array}
\right)\nonumber
\eeq
which renders the analysis of the extremization of (\ref{LagD}) in components a straightforward task. We obtain
\begin{eqnarray*}
d^\text{extended}(\mathbb{W},\bbL)=\frac{1}{2\rho_{11}^2 (\lambda_2^2-4 \lambda_1 \lambda_3) } \Big( -8 \lambda _3 \lambda _1 \text{Re}\left(w_1\right)-4 \lambda _2 \lambda _1
   \text{Re}\left(w_2\right)+4 \lambda _2^2 \,\text{Re}\left(w_1\right)-8
   \lambda _3 \lambda _1^2\\-2 \lambda _1 w_2 \left(w_2\right){}^*+\lambda _2
   w_2 \left(w_1\right){}^*-2 \lambda _3 w_1 \left(w_1\right){}^*+\lambda _2
   w_1 \left(w_2\right){}^* \Big) \,.
   \end{eqnarray*}

It is possible to minimize analytically the dual functional $d(\mathbb{W},\bbL)$ with respect to $\bbL$.
The resulting dual objective has been already shown in eq.~\eqref{dual bootstrap 12system}. 
Here we shall report the expressions for the critical amplitudes as function of $\mathbb{W}$ only.
   
In the extended unitarity region $4<s<(m_1+m_2)^2$ the critical amplitudes are given by
\begin{align}
M_{11\to 11}&=\frac{i}{\rho^2_{11}}\left(1+\frac{(W_2^2-4 W_1 W_3)^*}{|W_2^2-4 W_1 W_3|}\right),\nonumber\\ 
%\qquad \to \qquad S_{11}^{11}=-\frac{(W_2^2-4 W_1 W_3)^*}{|W_2^2-4 W_1 W_3|}
M_{11 \to 12}&=\frac{i}{2\rho_{11}^2}\frac{4 W_1^* W_2 W_3-|W_2|^2-|W_2^2-4 W_1 W_3|}{W_3 |W_2^2-4 W_1 W_3|},\label{extAp}\\
\IM M_{12 \to 12}&=\frac{|W_2|^4+|W_2|^2 |W_2^2-4 W_1 W_3|-4 W_3 \,\RE (W_2^2 W_1^*)}{4 \rho_{11}^2 |W_2^2-4 W_1 W_3| W_3^2}.\nonumber
\end{align}
%It should still be possible to proceed a bit further in matrix form in this case as well but for all practical purposes components work just as well. 

Notice that we cannot have direct access to $\RE M_{12\to 12}$, but we can reconstruct it from its imaginary part. This is of course related to the fact that our equations in the extended unitarity region, with $\rho$ non-invertible, are a bit more degenerate. 

In the unitarity region, $s>(m_1+m_2)^2$, the expressions of the critical amplitudes are much more involved.
It is convenient to introduce the two auxiliary functions
\begin{align}
\alpha&=\frac{1}{2\rho_{11}^2}\sqrt{\rho_{12}^4 |W_1|^2+\rho_{11}^4 |W_3|^2+\frac{1}{2}\rho_{11}^2\rho_{12}^2 (|W_2|^2+|W_2^2-4 W_1 W_3|)},\nonumber\\
\beta&=\frac{\alpha}{2i}\,\frac{4 W_2 W_1^* |W_3|^2+W_3 W_2^*(|W_2^2-4 W_1 W_3|-|W_2|^2)}{2 \IM W_2\, \RE W_2\, \RE(W_1 W_3)-\RE (W_2^2)\IM (W_1 W_3)}.\nonumber
\end{align}
The amplitudes can then compactly written as
\begin{align}
M_{11\to11}&=\frac{i}{2\alpha}\left(\frac{2\alpha +W_1^*}{\rho_{11}^2}-\frac{ W_3^* \beta}{\rho_{12}^2 \beta^*}\right),\nonumber\\
M_{11\to12}&=\frac{i}{\rho_{12}^2}\frac{W_3^*}{\beta^*},\label{regularAp}\\
M_{12\to12}&=-\frac{i}{2\alpha |\beta|^2 \rho_{12}^4}(W_3^*(4\alpha^2 \rho_{12}^2-|\beta|^2 \rho_{11}^2)+\rho_{12}^2 \beta^*(W_1^* \beta^* -2 \alpha(\beta+W_2^*))).\nonumber
\end{align}
Quite non-trivially, relations (\ref{extAp}) and (\ref{regularAp}) manifestly saturate extended and regular unitarity in our truncated space.

\section{Dual $\mathbb{Z}_2$ bootstrap}
\label{Z2appendix}
\subsection{Setup the primal problem}

Here we consider a simple application of the dual technology developed in Sec.~\ref{sec2} to the scattering of equal mass particles with different field parity: $1$ odd and $2$ even.
Defining the $S$-matrix element for the process $ij\to kl$ as $S_{ij}^{kl}= \,_{\text{out}}\langle kl | ij \rangle_{\text{in}}$, we can group the even and odd scattering processes into two $2\times 2$ matrices\footnote{Recall that in $1+1$ dimensions forward $12\to 12$ and backward $12\to 21$ scattering of non-identical particles are independent processes.}
\beq
\mathbb{S}^{\text{even}}=\begin{pmatrix}
S_{11 \to 11} & S_{11 \to 22} \\
S_{11 \to 22} & S_{22 \to 22}
\end{pmatrix}
\qquad 
\mathbb{S}^{\text{odd}}=\begin{pmatrix}
S_{12 \to 12} & S_{12 \to 21} \\
S_{12 \to 21} & S_{12 \to 12}
\end{pmatrix}.
\label{Sevenodd}
\eeq
Unitarity is simply given by the two positivity constraints
\beq
\mathbb{U}^{\text{even}}=\mathds{1}-\mathbb{S}^{\text{even}}( \mathbb{S}^{\text{even}})^\dagger \succeq 0, \qquad \mathbb{U}^{\text{odd}}=\mathds{1}-\mathbb{S}^{\text{odd}}( \mathbb{S}^{\text{odd}})^\dagger \succeq 0, \qquad s\geq4m^2.
\label{unitarityZ2}
\eeq
Analyticity and crossing properties are encoded into the dispersion relations
\beq
\mathcal{A}_a(s)=S_a(s)-S_a(\infty)+\frac{J g^2_a }{s-m^2}+\frac{J \mathcal{C}_{ab} g^2_b}{t(s)-m^2}-\frac{1}{\pi}\int_{4m^2}^\infty \left(\frac{\im S_a(z)}{z-s}+\frac{\mathcal{C}_{ab}\im S_b(z)}{z-t(s)}\right)dz=0,
\label{Z2disp}
\eeq
where $J=1/2\sqrt{m^2(4-m^2)}$ and $a,b=\{11\to11,22\to22,12\to12,11\to22,12\to21\}$. In this basis the crossing matrix is simply
\beq
\mathcal{C}=\begin{pmatrix}
{\color{darkblue}1} & 0 & 0 & 0 & 0 \\
0 & {\color{darkblue}1} & 0 & 0 & 0\\
0 & 0 & {\color{darkblue}1} & 0 & 0\\
0 & 0 & 0 & 0 & {\color{darkblue}1}\\
0 & 0 & 0 & {\color{darkblue}1} & 0
\end{pmatrix}.
\eeq
The processes $\{11\to 11,22\to 22,12\to12\}$ are invariant under crossing $s\to t=4m^2-s$. 
The last two processes $11\to22$ and $12\to 21$ are crossed of each other.

Because of $\mathbb{Z}_2$ symmetry there are only two independent couplings that we call $g_{112}$ and $g_{222}$.
They show up in the different processes as follows
\beq
\begin{array}{c|c|c}
\text{Amplitude} & \text{Exchange of particle } 1 & \text{Exchange of particle } 2  \\  \hline
11\to 11 &  0  &  {\color{blue} g_{112}^2} \\  \hline
22\to 22 & 0  & {\color{magenta} g_{222}^2 } \\ \hline
12\to 12 & {\color{blue} g_{112}^2} & 0\\ \hline
11\to 22 &  0  & {\color{blue} g_{112}} {\color{magenta} g_{222}} \\ \hline
12\to 21 &  {\color{blue} g_{112}^2}  & 0 \\ \hline
\end{array} \nonumber
\eeq

%\begin{align}
%&S_{11\to11}(s)=-J g^2_{112}\left(\frac{1}{s-m^2}+\frac{1}{4-s-m^2}\right)+\dots\nonumber\\
%&S_{22\to22}(s)=-J g^2_{222}\left(\frac{1}{s-m^2}+\frac{1}{4-s-m^2}\right)+\dots\nonumber\\
%&S_{12\to12}(s)=-J g^2_{112}\left(\frac{1}{s-m^2}+\frac{1}{4-s-m^2}\right)+\dots\nonumber\\
%&S_{11\to22}(s)=S_{12\to21}(4m^2-s)=-J \left(\frac{g_{112}g_{222}}{s-m^2}+\frac{g^2_{112}}{4-s-m^2}\right)+\dots,
%\end{align}
%with the jacobian $J=1/(2\sqrt{m^2(4-m^2)})$ the same for all the processes because of the equal masses.

One way to explore the space of allowed couplings is to formulate the problem in a radial form.
We define 
\beq
g_{112}=R \cos\theta,\qquad g_{222}=R\sin\theta,\nonumber
\eeq
and the vector $v(\theta)=\{\cos^2\theta,\sin^2\theta,\cos^2\theta, \sin\theta \cos\theta, \cos^2\theta\}$. Then for each fixed $\theta$ we solve:
\begin{mdframed}[frametitle={Primal $\mathbb{Z}_2$ Problem},frametitlealignment=\centering,
backgroundcolor=blue!6, leftmargin=0cm, rightmargin=0cm, topline=false,bottomline=false, leftline=false, rightline=false] 
\vspace{-0.5cm}
\begin{align} &\underset{\text{in } {\{R^2,S_a\}}}{\text{maximize}} &&  R^2\nonumber\\
& \text{constr. by}  &&\text{Res}_{m^2}(S_a)=v_a(\theta)R^2 \qquad\qquad \text{for } a=1,\dots,5\label{radialZ2}\\ 
&  && \mathcal{A}_a(s)=0\qquad\qquad\qquad\qquad\,\,\,\,\, \text{for } s>4m^2,\quad \text{for }a=1,\dots,5\nonumber\\
& &&\mathbb{U}^{\text{even}}(s) \succeq 0, \quad \mathbb{U}^{\text{odd}}(s) \succeq 0 \quad\,\,\, \text{for } s>4m^2.
 \label{primal bootstrap 11to11}
\end{align}
\end{mdframed}
This problem and therefore the space of the allowed couplings $\{g_{112},g_{222}\}$ has been already determined in~\cite{Paper4}.
Our aim is to give an equivalent dual formulation which makes the problem so simple that can be ran in few minutes using \texttt{Mathematica} on a standard laptop.

\subsection{Dual construction I: residue constraints}

As explained in Sec.~\ref{sec2}, the construction of the dual problem starts with the introduction of Lagrange multipliers for any constraint given in the primal problem~\eqref{primal bootstrap 11to11}.
The first set of linear constraints~\eqref{radialZ2} defines what we call ``radial problem'' -- see also~\cite{Monolith}. 
They can be easily taken into account introducing the Lagrangian
\beq
\mathcal{L}(R^2,S,\nu)=R^2+\sum_{a}\nu_a (\text{Res}_{m^2}(S_a)-v_a(\theta)R^2)=R^2\left(1-\sum_a \nu_a v_a(\theta)\right)+\sum_{a}\nu_a \text{Res}_{m^2}(S_a).\nonumber
\eeq
The Lagrange equation for $R^2$ yields simply the condition
\beq
1-\sum_a \nu_a v_a(\theta)=0,\nonumber
\eeq
and the problem can be cast in a simpler equivalent form
\begin{align}
&\min_{\nu_a}\quad && \biggr\{ \,  \max_{S_a} \qquad \sum_{a}\nu_a \text{Res}_{m^2}(S_a)\qquad  \text{constrained by }\nonumber\\
& && \qquad\qquad\qquad \mathcal{A}_a(s)=0, \qquad s\geq 4 m^2,\quad a=1,\dots,5\nonumber\\
& && \qquad\qquad\qquad\mathbb{U}^{\text{even}}(s) \succeq 0, \quad \mathbb{U}^{\text{odd}}(s) \succeq 0, \quad s\geq 4 m^2\, \biggr\}\nonumber\\
&  \text{constrained by } && 1-\sum_a \nu_a v_a(\theta)=0.
\label{Z2Dual1}
\end{align}

\subsection{Dual construction II: analyticity and crossing}

All crossing and analyticity properties of the various $S$-matrices involved in the $\mathbb{Z}_2$ system can be derived from the dispersion relations in eq.~\eqref{Z2disp}. 
Indeed, for each $s$ they can be viewed as a set of linear constraints enforcing a precise relation among the $\RE M(s)$ and $\IM M(s)$, otherwise independent.

For each component we introduce a dual scattering function $w_a$ and replace the objective in~\eqref{Z2Dual1} by
\beq
\mathcal{L}(S,\nu, w)=\sum_{a}\nu_a \text{Res}_{m^2}(S_a)+\sum_a \int_{4m^2}^\infty w_a(s) \mathcal{A}_a(s)\,ds.
\label{StartingLag2}
\eeq
The $w_a(s)$ are real functions in general. 
However, it is often useful to define analytic functions starting from $w_a(s)$ to simplify the analyticity and crossing constraint.
It can be shown that if we introduce a \emph{dual crossing} function $W_a(4m^2-s)=-\mathcal{C}_{ab} W_b(s)$ such that
\beq
W_a(s)=\frac{1}{\pi}\int_{4m^2}^\infty  \left(\frac{\im W_a(z)}{z-s}-\frac{\mathcal{C}_{ab}\im W_b(z)}{z-t(s)}\right)dz\nonumber
\eeq
and identify
\beq
w_a(s)=\im W_a(s), \qquad \text{for } s>4m^2,\nonumber
\eeq
the last term in eq.~\eqref{StartingLag2} becomes
\beq
\sum_a \int_{4m^2}^\infty w_a(s) \mathcal{A}_a(s)\,ds=\pi J \sum_a \text{Res}_{m^2}(S_a)W_a(m^2)+\sum_a \int_{4m^2}^\infty \im (W_a S_a)\,ds.
\label{Z2sumrule}
\eeq
%where $W_a(s)$ is a set of 5 analytic functions with normal cuts. Their boundary values are the lagrange multipliers associated to the dispersive constraints

Substituting eq.~\eqref{Z2sumrule} into the Lagrangian~\eqref{StartingLag2} allows us to maximize in the $\text{Res}_{m^2}(S_a)$ variables
\beq
\frac{\p}{\p \text{Res}_{m^2}(S_a)}\mathcal{L}(S,\nu,w)=\nu_a+\pi J W_a(m^2)=0, \qquad \text{for } a=1,\dots,5,\nonumber
\eeq
and use these 5 equations to eliminate the Lagrange multipliers setting $\nu_a=-\pi J W_a(m^2)$.
The radial constraint translates into a condition on the $W_a$ dual scattering functions
\begin{align}
&\min_{\nu_a,W_a}\quad && \biggr\{ \,  \max_{S_a} \qquad\sum_a \int_{4m^2}^\infty \im (W_a S_a)\,ds \qquad  \text{constrained by }\nonumber\\
& && \qquad\qquad\qquad\mathbb{U}^{\text{even}}(s) \succeq 0, \quad \mathbb{U}^{\text{odd}}(s) \succeq 0, \quad s\geq 4 m^2\, \biggr\}\nonumber\\
&  \text{constrained by } && 1+\pi J\sum_a v_a(\theta) W_a(m^2)=0.
\label{Z2Dual2}
\end{align}

\subsection{Dual construction III: unitarity}

The last constraint to add to the dual Lagrangian is unitarity.
This can be elegantly done if we cast the problem~\eqref{Z2Dual2} into a matrix form.
If we define the symmetric matrices 
\beq
\mathbb{W}^{\text{even}}=\begin{pmatrix}
2 W_1 & W_4\\
W_4 & 2 W_2
\end{pmatrix}, \qquad
\mathbb{W}^{\text{odd}}=\begin{pmatrix}
W_3 & W_5 \\
W_5 & W_3
\end{pmatrix}. \la{wmatrices}
\eeq
the objective of~\eqref{Z2Dual2} is just
\beq
\sum_a \int_{4m^2}^\infty \im (W_a S_a)\,ds = \int_{4m^2}^\infty \,ds\left( \frac{1}{2} \IM \left(\tr \mathbb{W}^{\text{even}} \mathbb{S}^{\text{even}}\right)+\frac{1}{2} \IM \left(\tr \mathbb{W}^{\text{odd}} \mathbb{S}^{\text{odd}}\right)\right),\nonumber
\eeq
recalling $\mathbb{S}^{\text{even/odd}}$ were introduced in~\eqref{Sevenodd}. 
It is then natural to introduce the semidefinite-positive matrix Lagrange multipliers $\bbL^{\text{even}}$ and $\bbL^{\text{odd}}$ 
\beq
\mathcal{L}(\mathbb{S},\mathbb{W},\Lambda){=}\int_{4m^2}^\infty ds\,\left(\frac{1}{2}\IM \left( \tr \mathbb{W}^{\text{even}} \mathbb{S}^{\text{even}} {+} \tr\mathbb{W}^{\text{odd}} \mathbb{S}^{\text{odd}}\right)+\frac{1}{2} \tr \bbL^{\text{even}} \mathbb{U}^{\text{even}}+\frac{1}{2} \tr \bbL^{\text{odd}} \mathbb{U}^{\text{odd}}\right),
\label{LagZ2Unitarity}
\eeq
So that 
\beq
\delta_{\mathbb{S}}\mathcal{L}{=}\int_{4m^2}^\infty ds\, \sum_{\eta=\text{even,odd}} \tr\!\left[\left(\frac{- \overline{\mathbb{W}}^{\eta}}{4i} -\frac{1}{2} \bbL^\eta \cdot \mathbb{S}^\eta \right) \delta\overline{\mathbb{S}}^\eta+\texttt{conjugate}   \right]
\label{LagZ2Unitarity}
\eeq
Since $0=\tr(\sigma_y \cdot \bar{\mathbb{S}})=\tr(\sigma_y \cdot {\mathbb{S}})=\tr(\sigma_y \cdot \delta\bar{\mathbb{S}})=\tr(\sigma_y \cdot \delta{\mathbb{S}})$ the parentheses does not need to vanish. It does need to be proportional to $\sigma_y$ with a proportionality constant which we can easily find by dotting it with the appropriate matrices: 
\beq
\frac{- \overline{\mathbb{W}}}{4i} -\frac{1}{2} \bbL \cdot \mathbb{S} = -\frac{1}{4i} \frac{\text{tr}(\bbL^{-1}\cdot \overline{\mathbb{W}}\cdot \sigma_y)}{\text{tr}(\bbL^{-1})} \sigma_y \label{new}
\eeq
where we dropped the implicit label $\eta$. For $\eta=\text{odd}$ this equation simplifies dramatically because\footnote{Notice $\mathbb{U}^{\text{odd}}$ is real and symmetric, we can take $\bbL^{\text{odd}}$ real and symmetric as well without loss of generality.}
\beq
\bbL^{\text{odd}}=
\begin{pmatrix}
\lambda_3 & \lambda_5 \\
\lambda_5 & \lambda_3.
\end{pmatrix}\nonumber
\eeq
and the right hand side of (\ref{new}) vanishes once we use (\ref{wmatrices}). In that case we therefore obtain
 the critical $S$-matrix in the odd sector as compactly given by $\mathbb{S}^{\text{odd}}=\frac{i}{4}(\bbL^{\text{odd}})^{-1}\overline{\mathbb{W}}^\text{odd}$. Furthermore, 
minimizing the dual functional over $\bbL$ is equivalent to solving the constraint equation $\mathds{1}=\mathbb{S}^{\text{odd}}(\mathbb{S}^{\text{odd}})^\dagger$ which determines\footnote{The square root of a matrix is not uniquely defined in general. Here we should pick the positive-semidefinite solution.}
%with $\bbL\succeq 0$\footnote{The square root of a matrix is not uniquely defined in general. Here we should pick the positive-semidefinite solution.}
\beq
%\mathds{1}=\mathbb{S}^{\text{odd}}(\mathbb{S}^{\text{odd}})^\dagger\qquad \to \qquad 
\bbL^{\text{odd}}=\frac{1}{2}\sqrt{\overline{\mathbb{W}}^\text{odd}\cdot {\mathbb{W}}^\text{odd}}.
\label{LambdaCritical}
\eeq
%
%\beq
%\mathbb{S}^{\text{odd}}=\frac{i}{4}(\bbL^{\text{odd}})^{-1}\overline{\mathbb{W}}^\text{odd}=\frac{i}{2}\left(\sqrt{\mathbb{W}^* \mathbb{W}}\right)^{-1}\mathbb{W}^*.
%\eeq
Plugging eq.~\eqref{LambdaCritical} into the dual functional we finally get
\beq
\inf_{\bbL} d(\mathbb{W},\bbL)^{\text{odd}}= D^{\text{odd}}(\mathbb{W})=\frac{1}{2} \int_{4m^2}^\infty \tr \left(\sqrt{\overline{\mathbb{W}}\cdot \mathbb{W}}\right)^{\text{odd}}ds.
\label{matrixlagrangian}
\eeq
In the even sector case such an honest explicit derivation is not available because of the very non-linear appearance of $\bbL$. Inspired by the simplicity of~\eqref{matrixlagrangian}
we guessed the matrix formulation of the dual problem
%\footnote{We have numerically checked that the guessed functional solves the lagrange equations.}
\begin{mdframed}[frametitle={Dual $\mathbb{Z}_2$ Problem},frametitlealignment=\centering,backgroundcolor=red!6, leftmargin=0cm, rightmargin=0cm, topline=false,bottomline=false, leftline=false, rightline=false] 
\vspace{-0.5cm}
\begin{align} &\underset{\text{in } {\{\mathbb{W}^{\text{even}},\mathbb{W}^{\text{odd}}\}}}{\text{minimize}} && \frac{1}{2}\int_{4m^2}^\infty\tr\!\!\left(\sqrt{\overline{\mathbb{W}}\cdot \mathbb{W}}\right)^{\text{odd}}+\tr\!\!\left(\sqrt{\overline{\mathbb{W}}\cdot \mathbb{W}}\right)^{\text{even}}ds,\nonumber\\
& \text{constr. by}  &&1+\pi J\sum_a v_a(\theta) W_a(m^2)=0.
 \label{dualZ2problem}
\end{align}
\end{mdframed}
This guess is correct. We checked it numerically and also derived it by brute force going to components. 

%We start with the odd sector which is simpler.
%One can show that the two conditions sufficient for unitarity in the odd sector are
%\beq
%\{1-|S_{12\to12}|^2-|S_{12\to 21}|^2 \geq 0, \RE (S_{12\to 12}S_{12\to 21}^*)=0  \} \implies \mathbb{U}^{\text{odd}}\succeq 0,
%\eeq
%corresponding to the usual diagonal unitarity inequality and to the off-diagonal equality condition.
%{\color{red} Cleaning up the logic, this condition is not necessary. I will comment about the possibility of perform an sdpb search and so on somewhere}
%
%As we have learned, we can introduce a lagrange multiplier function for each constraint
%\begin{align}
%&\mathcal{L}(S,\nu,W,\lambda)= \sum_a \int_4^\infty \im (W_a S_a)\,ds\nonumber\\
%&+ \int_4^\infty \lambda_1 (1-|S_{12\to12}|^2-|S_{12\to 21}|^2)\,ds+\int_4^\infty \lambda_2 \RE (S_{12\to 12}S_{12\to 21}^*)\,ds,
%\end{align}
%with $\lambda_1 \succeq 0$.
%
%In the even sector the conditions are four 
%\begin{align}
%&\{1-|S_{11\to11}|^2-|S_{11\to 22}|^2 \geq 0,1-|S_{11\to22}|^2-|S_{22\to 22}|^2 \geq 0, \nonumber\\
%&\RE(S_{11\to 11}S_{11\to 22}^*+S_{11\to 22}S_{22\to 22}^*)=0, \IM(S_{11\to 11}S_{11\to 22}^*+S_{11\to 22}S_{22\to 22}^*)=0  \} {\implies} \mathbb{U}^{\text{even}}\succeq 0
%\end{align}

\begin{figure}[ht]
	\centering 
	\includegraphics[width=\linewidth]{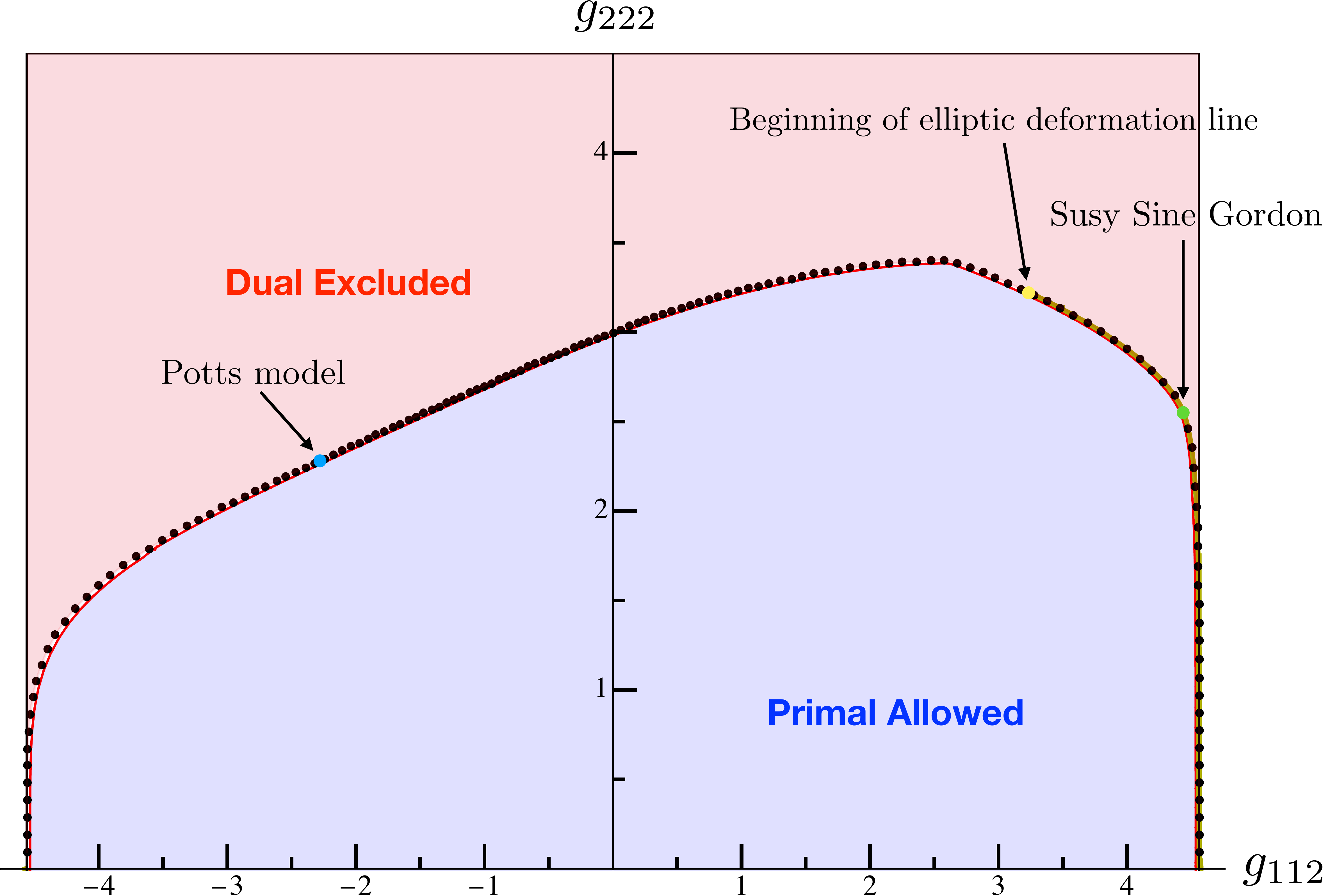}
	\caption{Space of the $\mathbb{Z}_2$ symmetric coupling constants $\{g_{222}, g_{112}\}$. We restrict to the UHP due to $g \leftrightarrow -g$ symmetry. The black dots are obtained minimizing the objective in~\eqref{explicitZ2objective}. The red solid line was obtained in~\cite{Paper4} running the primal problem. The dual data have been obtained with very little effort: in this plot $N_{\text{max}}=5$ for all dual scattering functions ansatz. The blue and green dots mark respectively the well known integrable $3$-states Potts and supersymmetric Sine-Gordon. We recall that starting at $\theta=\pi/4$, the yellow dot, and all the way to $\theta=\pi/2$ the S-matrix saturating the boundary is known analytically and correspond to the elliptic deformation of supersymmetric Sine-Gordon -- see~\cite{Paper4, ToAppearSUSY} for details.}
	\label{figZ2}
\end{figure}

Despite the simplicity of the matrix formulation, it is convenient to go back to components when performing numerical explorations.
The matrix $\overline{\mathbb{W}}\cdot \mathbb{W}$ for both odd and even sectors is a $2\times 2$ hermitian matrix. We can therefore use the general formula
\beq
M=\begin{pmatrix}
A & B \\
B^* & D
\end{pmatrix} \qquad \to \qquad
\tr \sqrt{M}=\sqrt{A+D+2\sqrt{A D-|B|^2}}\nonumber
\eeq
to derive the explicit form of the functional.
Applying this formula we get pretty straightforwardly that the first line in (\ref{dualZ2problem}) is given by 
\begin{align}
%& \frac{1}{2}\int_{4m^2}^\infty\tr\left(\sqrt{\mathbb{W}^\dagger \mathbb{W}}\right)^{\text{odd}}+\tr\left(\sqrt{\mathbb{W}^\dagger \mathbb{W}}\right)^{\text{even}}ds=\nonumber\\
\frac{1}{\sqrt{2}}\int\limits_{4m^2}^\infty\!\! ds\,\Big(\sqrt{|W_3|^2{+}|W_5|^2{+}|W_3^2{-}W_5^2|} +\!\sqrt{|W_4|^2{+}2|W_1|^2{+}2|W_2|^2{+}|W_4^2{-}4 W_1 W_2|}\,\Big) \,,
\label{explicitZ2objective}
\end{align}
which is the objective we minimize in practice. 

%\begin{align}
%\min_{\nu_a, W_a} \frac{1}{\sqrt{2}}\int_4^\infty ds\, &\sqrt{|W_3|^2+|W_5|^2+\sqrt{|W_3|^2+|W_5|^2-4 \RE(W_3 W_5^*)}}\nonumber\\
%&+\sqrt{|W_4|^2+2(|W_1|^2+|W_2|^2)+|W_4^2-4 W_1 W_2|}\nonumber\\
% \text{subject to } \qquad &1-\sum_a \nu_a v_a(\theta)=0, \qquad  \nu_a+\pi J W_a(m^2)=0.
%\end{align}
%\begin{figure}[ht]
%	\centering 
%	\includegraphics[width=\linewidth]{Z2plotv2-crop.pdf}
%	\caption{Space of the $\mathbb{Z}_2$ symmetric coupling constants $\{g_{222}, g_{112}\}$. The black dots are obtained minimizing the objective in~\eqref{explicitZ2objective}. The red solid line was obtained in~\cite{Paper4} running the primal problem. The dual data have been obtained with very little effort: in this plot $N_{\text{max}}=5$ for all dual scattering functions ansatz. The blue and green dots mark respectively the well known integrable $3$-states Potts and supersymmetric Sine-Gordon respectively. We recall that starting at $\theta=\pi/4$ and all the way to $\theta=\pi/2$ the S-matrix saturating the boundary is known analytically and correspond to the elliptic deformation of supersymmetric Sine-Gordon -- see~\cite{Paper4, ToAppearSUSY} for details.}
%	\label{figZ2}
%\end{figure}

\subsection{Dual problem numerics}

For the dual scattering functions associated to the crossing invariant processes $11\to11, 22\to22$, and $12\to12$ we can simply write the following anti-crossing ansatz
\beq
W_a(s)=\frac{1}{\sqrt{s(4m^2-s)}}(\rho(s)-\rho(t(s)))\left( \sum_{n=0}^{N_{\text{max}}^{(a)}} \alpha_n^{(a)} (\rho(s)^n+\rho(t(s))^n) \right),\quad a=1,2,3.\nonumber
\eeq
For the objective functional~\eqref{explicitZ2objective} we must require that $W_a\sim s^{-3/2}$ for $s\to\infty$. Our ansatz for these components has automatically the right behavior
since $\rho(s)-\rho(t)\sim s^{-1/2}$.

We can package the $11\to 22$ and $12\to 12$-backward into a single scattering function not symmetric under crossing
\beq\nonumber
W_4(s)=\frac{1}{\sqrt{s(4m^2-s)}}\left(\sum_{n=0}^{N_{\text{max}}^{(4)}} \beta^{(1)}_n \rho(s)^n+\beta^{(2)}_n \rho(t(s))^n \right).
\eeq
However, such an ansatz does not automatically decay with the right power and one must tune one of the free parameters.
The numerical results for the dual radial problem~\eqref{dualZ2problem} are shown in figure~\ref{figZ2}.

%{\color{magenta} Remove:} In the unitarity region, $s>(m_1+m_2)^2$, the expressions of the critical amplitudes are much more involved.
%It is convenient to introduce the two auxiliary functions
%\begin{align}
%\alpha&=\frac{1}{2\rho_{11}^2}\sqrt{\rho_{12}^4 |W_1|^2+\rho_{11}^4 |W_3|^2+\frac{1}{2}\rho_{11}^2\rho_{12}^2 (|W_2|^2+|W_2^2-4 W_1 W_3|)},\\
%\beta&=\frac{\alpha}{2i}\,\frac{4 W_2 W_1^* |W_3|^2+W_3 W_2^*(|W_2^2-4 W_1 W_3|-|W_2|^2)}{2 \IM W_2\, \RE W_2\, \RE(W_1 W_3)-\RE (W_2^2)\IM (W_1 W_3)}.
%\end{align}
%The amplitudes can then compactly written as
%\begin{align}
%M_{11\to11}&=\frac{i}{2\alpha}\left(\frac{2\alpha +W_1^*}{\rho_{11}^2}-\frac{ W_3^* \beta}{\rho_{12}^2 \beta^*}\right),\\
%M_{11\to12}&=\frac{i}{\rho_{12}^2}\frac{W_3^*}{\beta^*},\\
%M_{12\to12}&=-\frac{i}{2\alpha |\beta|^2 \rho_{12}^4}(W_3^*(4\alpha^2 \rho_{12}^2-|\beta|^2 \rho_{11}^2)+\rho_{12}^2 \beta^*(W_1^* \beta^* -2 \alpha(\beta+W_2^*))).
%\end{align}
%
%

\end{document}